\documentclass[final,3p,times,twocolumn]{elsarticle}
\usepackage{amssymb}
\usepackage{url}
\usepackage{graphicx}
\usepackage{subfig}

\journal{Nuclear Instruments and Methods A}

\begin{document}

\begin{frontmatter}



\title{Performance of the diamond-based beam-loss monitor system of Belle II}


\author[Cracow]{S.~Bacher}
\author[UniTS,SNS]{G.~Bassi}
\author[INFN]{L.~Bosisio}
\author[Elettra,INFN]{G.~Cautero}
\author[INFN]{P.~Cristaudo}
\author[INFN]{M.~Dorigo}
\author[UniTS,INFN]{A.~Gabrielli}
\author[Elettra,INFN]{D.~Giuressi}
\author[KEK]{K.~Hara}
\author[INFN]{Y.~Jin}
\author[UniTS,INFN,IPMU]{C.~La~Licata}
\author[INFN]{L.~Lanceri}
\author[UniTS,INFN]{R.~Manfredi}
\author[KEK]{H.~Nakayama}
\author[KEK]{K.~R.~Nakamura}
\author[Hawaii]{A.~Natochii}
\author[Pisa]{A.~Paladino}
\author[Pisa]{G.~Rizzo}
\author[UniTS,INFN]{L.~Vitale}
\author[Vienna]{H.~Yin}

\address[UniTS]{Dipartimento di Fisica, Universit\`a di Trieste, I-34127 Trieste, Italy}
\address[INFN]{INFN, Sezione di Trieste, I-34127 Trieste, Italy}
\address[Elettra]{Elettra Sincrotrone Trieste SCpA, AREA Science Park, I-34149 Trieste, Italy}
\address[SNS]{now at: Scuola Normale Superiore, I-56126 Pisa, Italy}
\address[IPMU]{now at: Kavli Institute for the Physics and Mathematics of the Universe (WPI), University of Tokyo, Kashiwa 277-8583, Japan}
\address[Pisa]{INFN Sezione di Pisa and Dipartimento di Fisica, Universit\`a di Pisa, I-56127 Pisa, Italy}
\address[KEK]{High Energy Accelerator Research Organization (KEK), Tsukuba 305-0801, Japan}
\address[Cracow]{H. Niewodniczanski Institute of Nuclear Physics, Krakow 31-342, Poland}
\address[Vienna]{Institute of High Energy Physics, Vienna 1050, Austria}
\address[Hawaii]{University of Hawaii, Honolulu, Hawaii 96822, USA}

\begin{abstract}
We designed, constructed and have been operating a system based on single-crystal synthetic diamond sensors, to monitor the beam losses at the interaction region of the SuperKEKB asymmetric-energy electron-positron collider. The system records the radiation dose-rates in positions close to the inner detectors of the Belle II experiment, and protects both the detector and accelerator components against destructive beam losses, by participating in the beam-abort system. It also provides complementary information for the dedicated studies of beam-related backgrounds. We describe the performance of the system during the commissioning of the accelerator and during the first physics data taking.
\end{abstract}

\begin{keyword}
sCVD diamond sensor \sep beam-loss monitoring \sep accelerator interlocks


\end{keyword}

\end{frontmatter}

%\linenumbers

\section{Introduction}
\label{sec:introduction}
Beam-loss monitoring is an important component of every particle-accelerator system. For the SuperKEKB~\cite{Ohnishi:2013fma} electron-positron collider and the Belle II experiment~\cite{Abe:2010gxa,Adachi:2018qme} this function is crucial, specially in the interaction region, since the accelerator is aiming at an unprecedented design luminosity of $8 \times 10^{35}$~cm$^{-2}$s$^{-1}$~\cite{Ohnishi:2013fma} with highly-focused intense beams. The detector has a projected lifetime of at least a decade, to collect a very large sample of electron-positron annihilation events, and explore extensions of the Standard Model of fundamental interactions.

The inner Belle II detectors are silicon-based pixel and microstrip sensors with integrated front-end electronics. The Belle II silicon technology is designed to sustain a dose of $10$ to $20$~Mrad, that is $0.1 - 0.2$~MGy, before significant performance degradation. It is therefore necessary to continuously monitor the dose rates from beam losses, and keep the radiation budget under control while the experiment integrates the planned luminosity. %We use $1$~rad~$= 10^{-2}$~Gy as dose unit in the following.

High dose rates, absorbed in short time intervals, may cause irreversible local damage to essential parts of the inner detector, and disrupt the accelerator operation by inducing quenches in the neighboring superconducting magnets of the final-focus system. A beam-loss monitor in this region must therefore be able to deliver beam-abort triggers to the accelerator control system, with a latency comparable with the revolution period of the accelerator rings.

In addition, knowledge of the intensity and timing patterns of beam losses in the interaction region offers essential information to optimize the collider operations and increase luminosity. 

To meet these challenges we designed, built, and have been operating a radiation monitoring system based on single-crystal diamond sensors, grown by chemical vapour deposition (sCVD sensors), and read-out by custom-designed electronics. This paper describes the diamond system requirements, construction, and its early operations and performance.

\section{Belle II interaction region and vertex detector}
\label{sec:B2_IR}

\begin{figure}[htbp]
	\begin{center}
		\includegraphics[width=7.0cm]{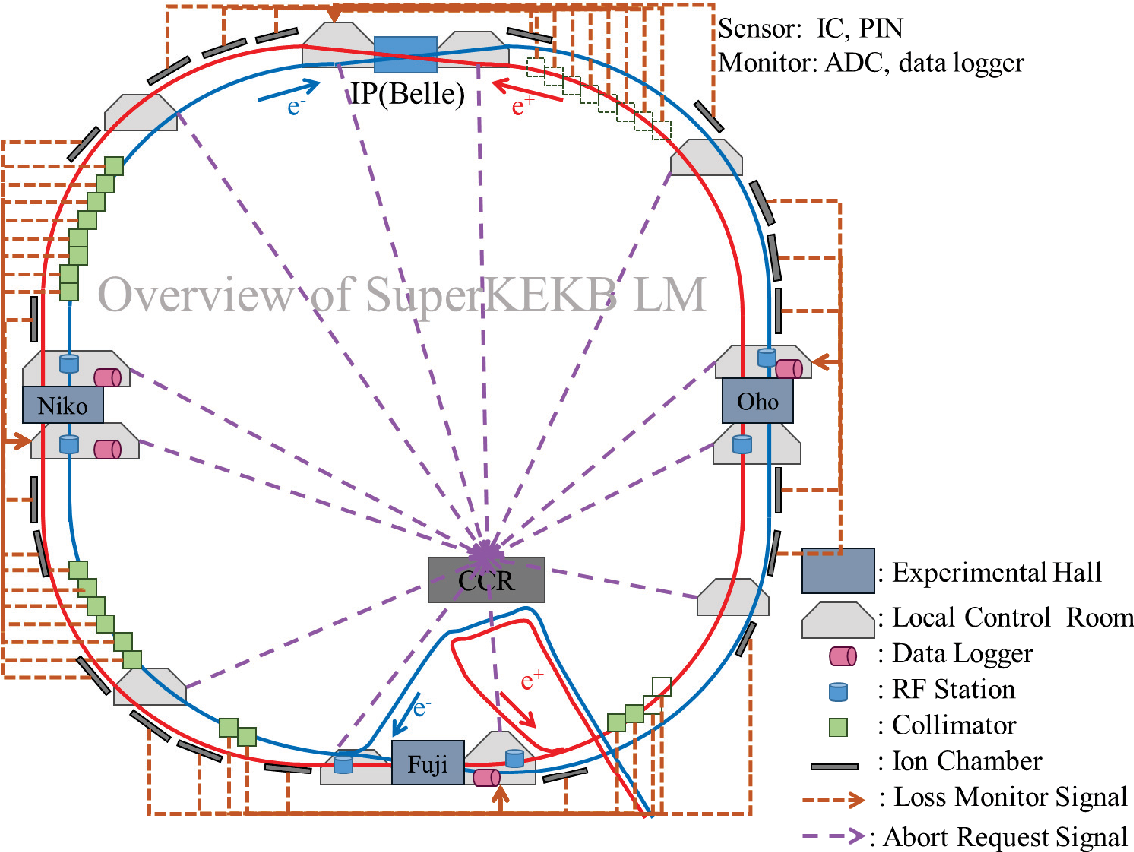}
		\caption{The SuperKEKB beam loss monitor and abort system, from reference~\cite{Ikeda:2017xtn}.}
		\label{fig:SKB_BLM}
	\end{center}
\end{figure}

The SuperKEKB layout is sketched in Figure~\ref{fig:SKB_BLM}~\cite{Ikeda:2017xtn} with some features of the beam abort system. Information from hardware components such as the RF, vacuum, magnets systems and the beam loss monitor signals are collected at local control rooms (LCR); their abort requests, signalling abnormal accelerator and beam conditions, are sent to the central control room (CCR) to trigger the abort kicker magnets. 

The Belle II detector~\cite{Abe:2010gxa,Adachi:2018qme} is a spectrometer surrounding the intersection of the $7$~GeV ``high-energy" electron ring (HER) and $4$~GeV ``low-energy" positron ring (LER). The inner silicon-based vertex detector (VXD) has two layers of pixel detectors (PXD) directly mounted on the beam pipe, and four layers of microstrip  silicon vertex detector (SVD) ladders supported by the cones and rings sketched in Figure~\ref{fig:RadView}. The superconducting final-focus magnets (QCS) of SuperKEKB extend partly inside the spectrometer, close to the beam-pipe bellows of the intersecting rings.

%The diamond-based monitor is an essential component of the beam loss and abort system, protecting both the Belle II vertex detector and the QCS magnets in the interaction region.

\section{The diamond-based beam-loss monitor system}
\label{sec:diamond_system}

The commissioning and data taking phases of SuperKEKB and Belle II are summarized in Table~\ref{tab:Phase}.

\begin{table}
\centering
\caption{
The three running phases of SuperKEKB and Belle II.
}
\label{tab:Phase} 
\begin{tabular}{ll}
\hline\noalign{\smallskip}
Phase/Year & Collisions / Detector \\
\noalign{\smallskip}\hline\noalign{\smallskip}
{\small 1 / 2016} & {\small No collisions / BEAST II~\cite{Lewis:2018ayu} }\\
{\small 2 / 2018} & {\small Collisions / BEAST II $\&$ Belle II, no VXD }\\
{\small 3 / 2019 $-$} & {\small Collisions / Full Belle II with VXD }\\
\hline\noalign{\smallskip}
\noalign{\smallskip}
\end{tabular}
\end{table}

\begin{figure*}
	\begin{center}
		\includegraphics[width=11cm]{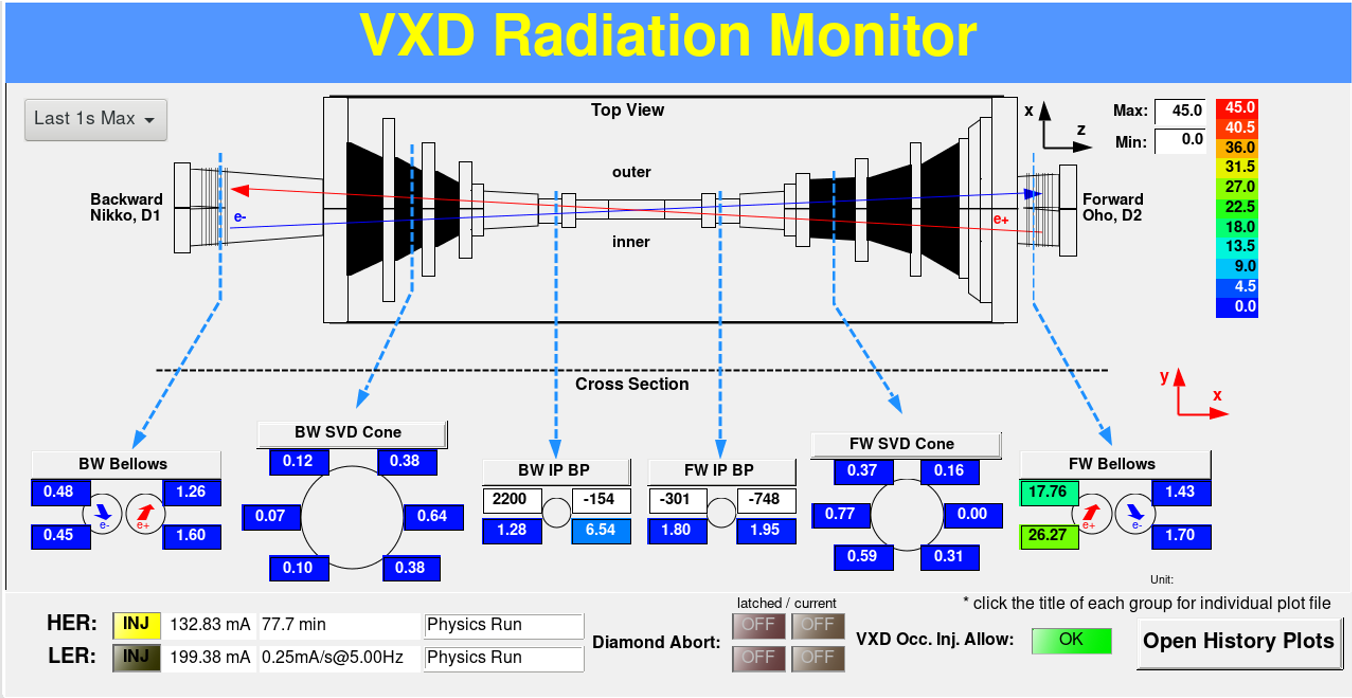}
		\caption{Sketch of the location of the $28$ diamond detectors on the beam pipe, the SVD support cones, and the beam-pipe bellows, as shown by the on-line display of dose rates, available to Belle II and SuperKEKB operators.}
		\label{fig:RadView}
	\end{center}
\end{figure*}

\begin{figure*}
	\begin{center}
		\includegraphics[width=15cm]{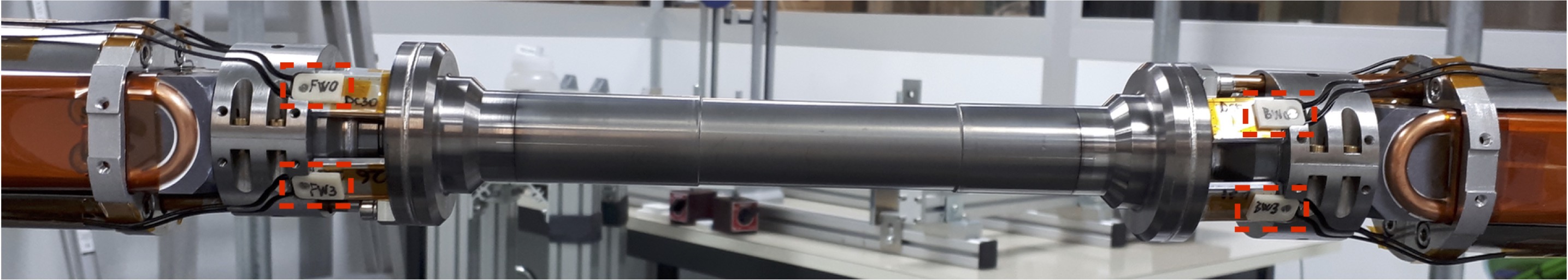}      \\
		\includegraphics[width=8.8cm]{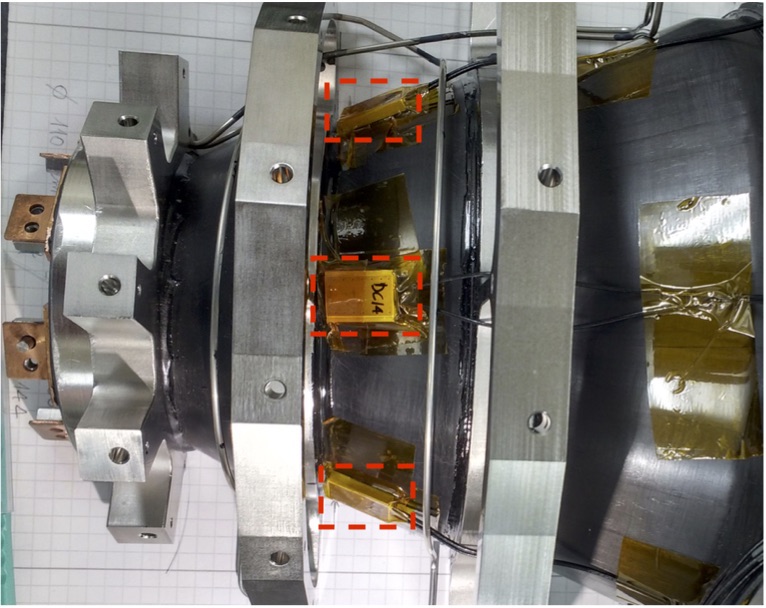}
		\includegraphics[width=6.2cm]{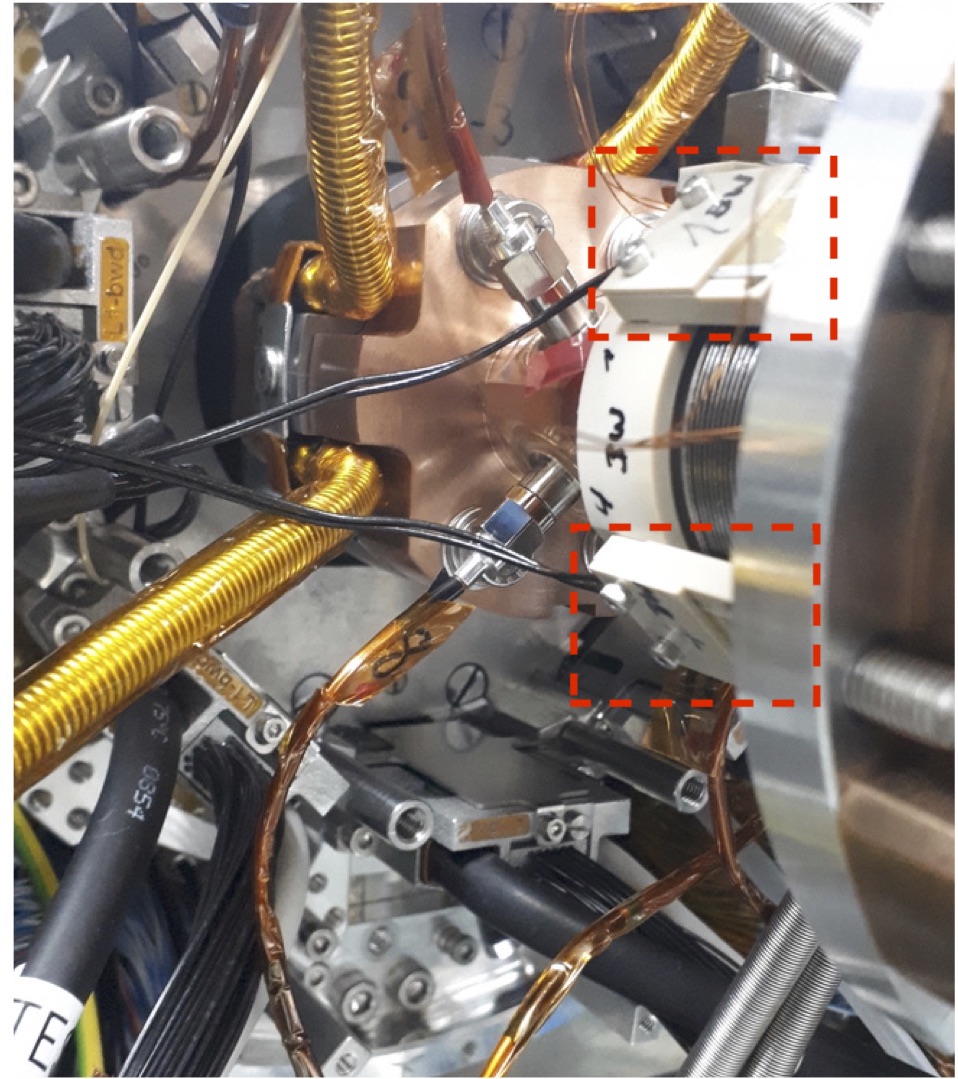}
	\caption{Photographs of diamond detectors (red dashed boxes) mounted on the beam pipe (top), on the backward SVD support cone (bottom left), and on the bellows close to the backward QCS (bottom right).}
	\label{fig:diamonds_location}
	\end{center}
\end{figure*}

Four prototype diamond detectors and their readout were installed and operated during the initial accelerator commissioning Phase 1 in 2016, as part of the BEAST~\cite{Lewis:2018ayu} project for initial beam-background studies; eight detectors were installed on the final beam pipe for the Phase 2 commissioning run, with first collisions and a provisional inner detector installed, in 2018. The final system consists of $28$ diamond detectors, mounted on the beam pipe (BP) near the interaction point (IP), on the SVD support cones, and on the beam-pipe bellows, close to the QCS magnets (Figures~\ref{fig:RadView} and~\ref{fig:diamonds_location}). We installed it in 2018 and have been operating it since then, for the physics data taking (Phase 3) with the complete Belle II apparatus.

\subsection{System requirements}
\label{subsec:system_requirements}

The KEKB peak luminosity was exceeded and a new world record of $2.4 \times 10^{34}$~cm$^{-2}$s$^{-1}$ was established by SuperKEKB during Phase 3 in June 2020. It was achieved with a product of beam currents that was less than 25\% that of KEKB, and by the initial implementation of the “nano-beam scheme” and the ``crab waist scheme''~\cite{Ohnishi:2020,Raimondi:2008zzb}: the vertical height of the beams at the interaction point was squeezed to about 220 nanometers. To approach the design luminosity, almost a factor $40$ higher, the beam height at the IP will have to decrease to approximately 50 nanometers, and the product of beam currents increase to four times that of KEKB~\cite{Ohnishi:2013fma}.

This luminosity enhancement is associated with harsher background conditions. The main beam-background sources are Touschek scattering, radiative Bhabha scattering, electron-positron pair production in photon-photon scattering, and off-momentum particles from beam-gas interactions~\cite{Lewis:2018ayu}. The intensities of these backgrounds are strongly dependent on the beam optics. At the design luminosity, QED backgrounds will become relevant.

We based the initial requirements of the monitoring system on background simulations and on the extrapolation of the experience of KEKB~\cite{Arinaga:2013pxa}, the BaBar experiment at PEP-II (SLAC)~\cite{Re:2004rc,Edwards:2005hi}, CDF at the Tevatron (FermiLab)~\cite{Sfyrla:2007ng}, ATLAS~\cite{ref_atlas} and CMS~\cite{ref_cms} at LHC. To design the read-out electronics, we assumed a sensor current to dose-rate conversion factor in the range $0.1 - 1$~(mrad/s)/nA; after calibrations, our sCVD sensors turned out to have sensitivities larger by an order of magnitude, of about $35$~(mrad/s)/nA on average.

We expected to amplify and digitize diamond currents from pA up to several mA, to be sensitive to both very low beam losses with radiation levels below $1$~mrad/s, and large spiky losses due to noisy beam injections or machine operation accidents, of the order of $10$~krad/s or more.

The time response, range, and precision of the associated read-out electronics were initially constrained as follows.
\begin{itemize}
    \item The time scale relevant for beam abort is the $10$~$\mu$s revolution time of electrons and positrons in the SuperKEKB rings: the sampling and comparison of signals with abort thresholds must match the corresponding frequency of $100$~kHz, or better.
    \item Localised radiation damage in SVD microstrip detectors may occur when a radiation dose in excess of O($1$~rad) is delivered in a time interval of O($1$~ms)~\cite{Edwards:2005hi}. To produce ``fast" abort requests, the digitised diamond current is integrated (averaged) in moving sums, updated and compared with an abort threshold every $10$~$\mu$s. With an integration time of about $1$~ms, a precision at the level of about $10$~nA is required on a $5$~mA measurement range for such aborts. The integration time and thresholds must be programmable, to satisfy evolving requirements.
    \item Finally a recording rate of $10$~Hz is chosen for dose rate monitoring, archiving and integration. The sensitivity obtained by averaging the dose rates on these time scales is also available for ``slow" abort requests, that could be delivered to the accelerator by the slow control software rather than by the diamond electronics directly.
\end{itemize}

\subsection{Diamond detectors}
\label{subsec:diamond_detectors}
Diamond sensors, with a bias voltage applied on two opposite electrodes, operate functionally as solid-state ionisation chambers, delivering a current induced by the drift of electrons and holes produced by ionisation in the diamond bulk. Their compact size and radiation hardness are well-suited for radiation monitoring in the interaction region; the temperature-independent response allows operation without additional temperature monitoring and related corrections.

The assembly, test, and calibration of the $28$ sCVD diamond detectors are detailed in ref.~\cite{Bassi:2020xy}.

\begin{figure}
	\centering
		\includegraphics[width=5cm]{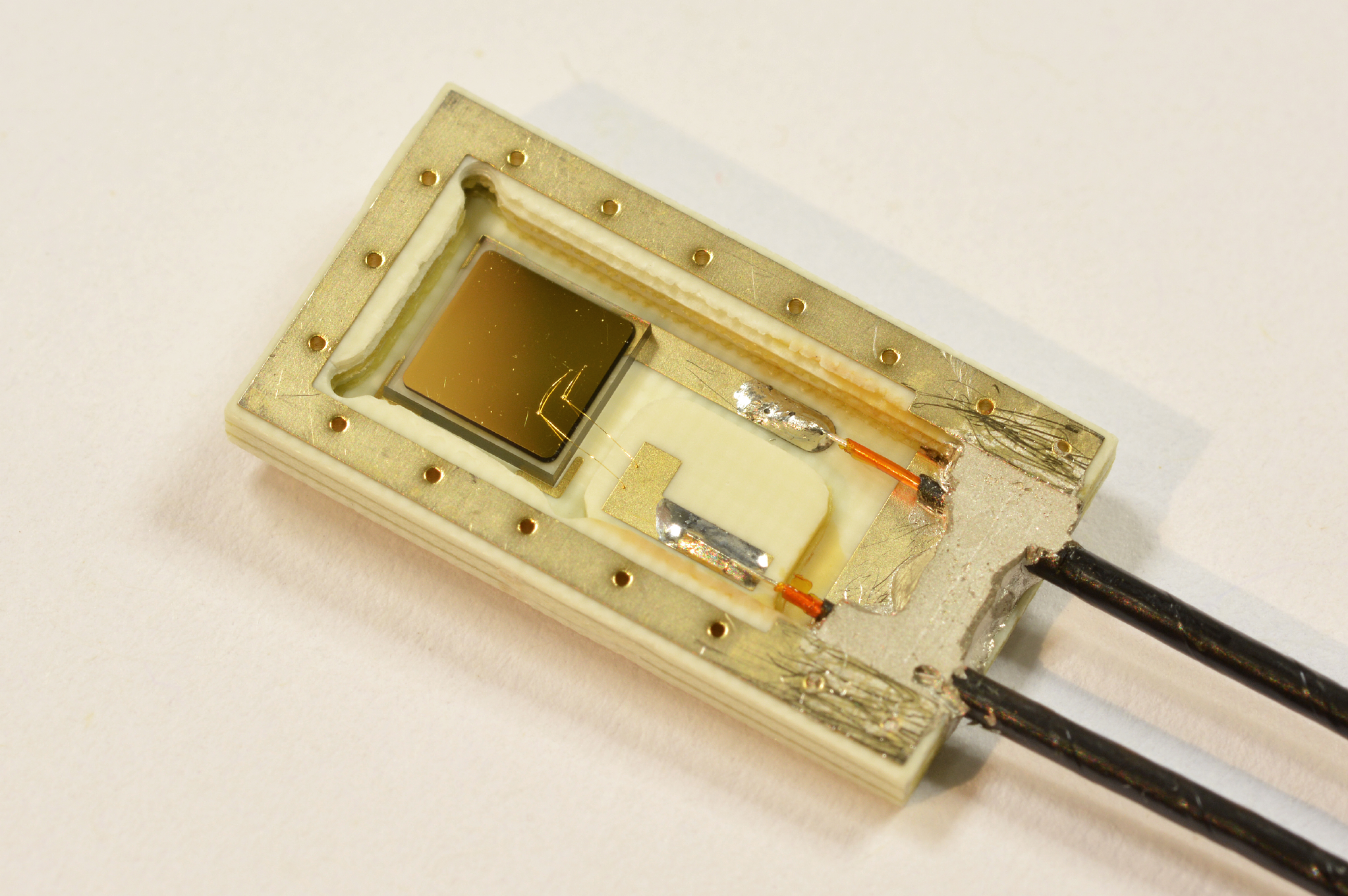}
	\caption{Diamond sensor packaged into a detector unit. The sensor is glued on a Rogers printed-circuit board; electrical contact is established between the two electrodes (front/back) and the inner conductors of two miniature coaxial cables; outer shielding is also provided.}
	\label{fig:diamonds_package}
\end{figure}

The electronic grade sCVD sensors were grown by Element Six~\cite{ref_e6} in the standard size $4.5 \times 4.5 \times 0.5$~mm$^{3}$;  two (Ti + Pt + Au), $(100 + 120 + 250)$~nm electrodes were deposited on both faces by CIVIDEC~\cite{ref_cividec}. We mounted each sensor on a small Rogers~\cite{ref_rogers} printed circuit board, providing both the mechanical support and the electrical screening, completed by a thin aluminum cover. The inner conductors of two thin coaxial cables were soldered to two pads, connected to the sensor electrodes by conductive glue (back side) and gold wire ball bonding respectively. The outer conductors of the cables were connected together and with the external detector shield, as shown in Figure~\ref{fig:diamonds_package}.

Each detector was characterized by measuring~\cite{Bassi:2020xy}
\begin{itemize}
   \item the current $I$ as a function of the bias voltage $V$, in the dark and in the presence of ionization by a $\beta$ radioactive source;
   \item the stability of the current signal in time;
   \item the transport properties of charge carriers, electrons and holes, using a $5$~kBq $^{241}$Am source of monochromatic $\alpha$-particles.
   \item a current to dose-rate calibration factor $k$, using an almost point like $3$~MBq $^{90}$Sr radioactive $\beta$ source located at varying distance.
\end{itemize}

The dose rate, expressed in mrad/s, is proportional to the measured current $I_{m}$ (nA),
\begin{eqnarray}
\label{eq:doserate_current}
\frac{dD}{dt} & = & \frac{1}{m} \frac{dE}{dt} = \frac{1}{m} \frac{I_{m}}{G} \frac{E_{eh}}{q_{e}} = \frac{F}{G}I_{m} = k I_{m}\\
F & = & \frac{E_{eh}}{m q_{e}} ,
\end{eqnarray}
where $m$ is the sensor mass, $E_{eh}$ the average ionization energy per electron-hole pair, $q_{e}$ the elementary charge. The unit conversion factor $F$ is computed as $F =  34.9$~(mrad/s)/nA. A bias voltage of $100$~V is chosen for full charge-collection efficiency. For a typical measured current of $1$~nA, the dose rate is $34.9$~mrad/s if $G = 1$.

The dimensionless factors $G$ describe possible deviations of each diamond sensor from the ideal behaviour in stationary conditions: 100\% efficiency in the collection of charge carriers from the assumed active volume, zero net trapping-detrapping rate, and absence of charge injection from the electrodes.
They are obtained from the ratios of the measured current to the corresponding energy deposit per unit time and sensor mass, estimated by a detailed simulation of the source and of the set-up. Similar measurements and simulations on a reference silicon diode are used to account for the source activity and to reduce the uncertainties of the set-up model. The measured $G$ values cluster around $G \simeq 1$, with maximum variations of about $50\%$. 

The uncertainty in the determination of the calibration factors $k$ is dominated by systematic effects; we estimate an accuracy of $8\%$~\cite{Bassi:2020xy}. 
The linear behaviour (equation~\ref{eq:doserate_current}) is confirmed by measurements performed on a sub-sample of diamond detectors with a $^{60}$Co $\gamma$ source at dose rates and currents larger by two orders of magnitude.

\subsection{Electronics}
\label{subsec:electronics}

\begin{figure}[htbp]
\begin{center}
\includegraphics[width=6cm]{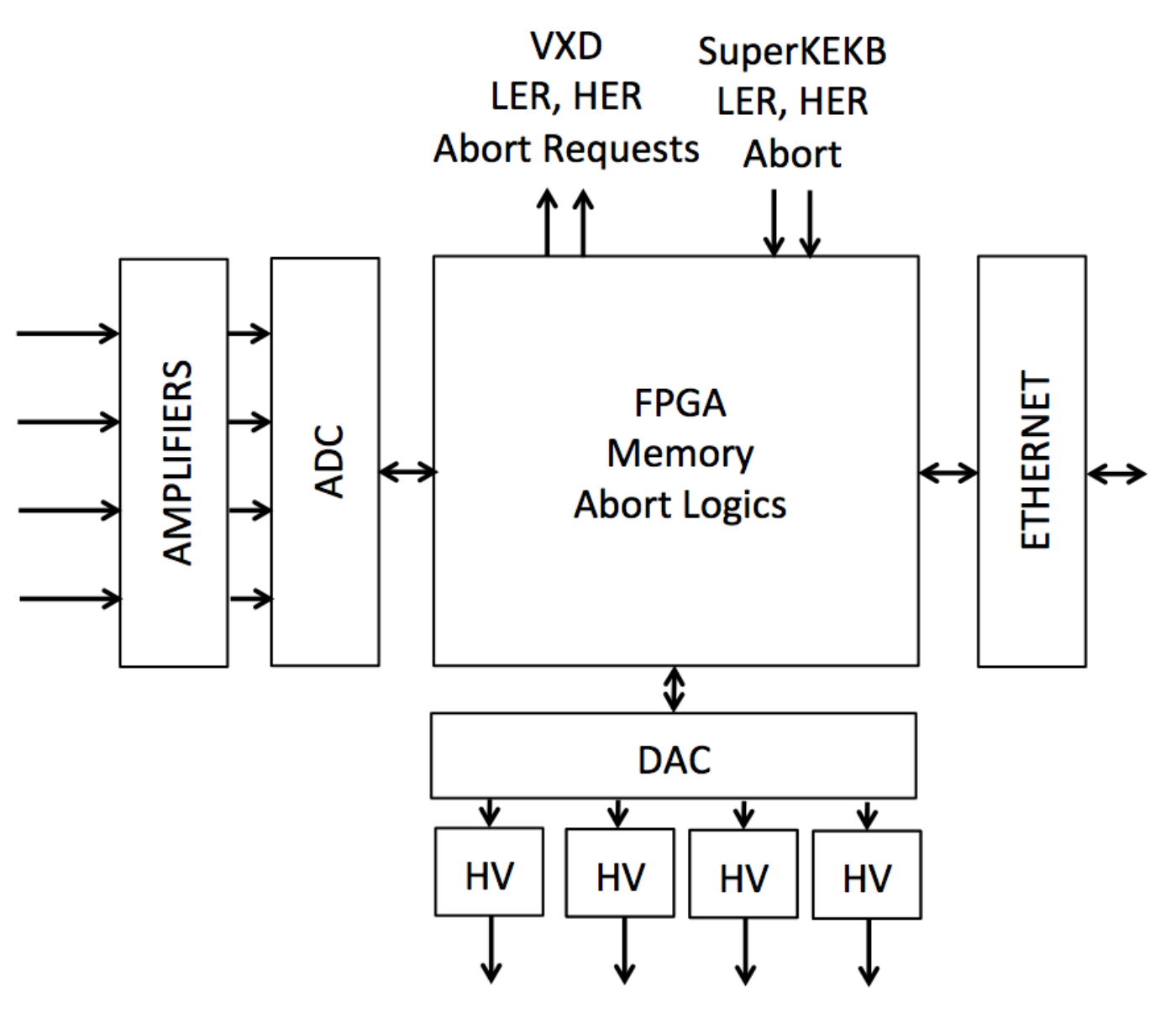}
\caption{Diamond control unit (DCU) block diagram. Arrows on the left side indicate currents from diamond sensors and arrows on the down side indicate high voltage applied on the diamond sensors.}
\label{fig:DCU}
\end{center}
\end{figure}

The $28$ diamond detectors are controlled by seven diamond control units (DCU). 
Each DCU pilots four diamond detectors, as sketched in the block diagram of Figure~\ref{fig:DCU} .

The digital core is a board hosting a Cyclone V FPGA by Intel~\cite{ref_cyclone}, which receives commands via an Ethernet interface, drives four HV modules independently through a DAC, and accepts input data from an analog module with amplifiers and ADC conversion, as sketched in Figure~\ref{fig:DCU}. 

The DCU is also able to deliver VXD abort requests separately for the electron higher-energy ring (HER) and the positron lower-energy ring (LER), and receives the SuperKEKB abort signals.

\begin{figure}[htbp]
\begin{center}
\includegraphics[width=7.5cm]{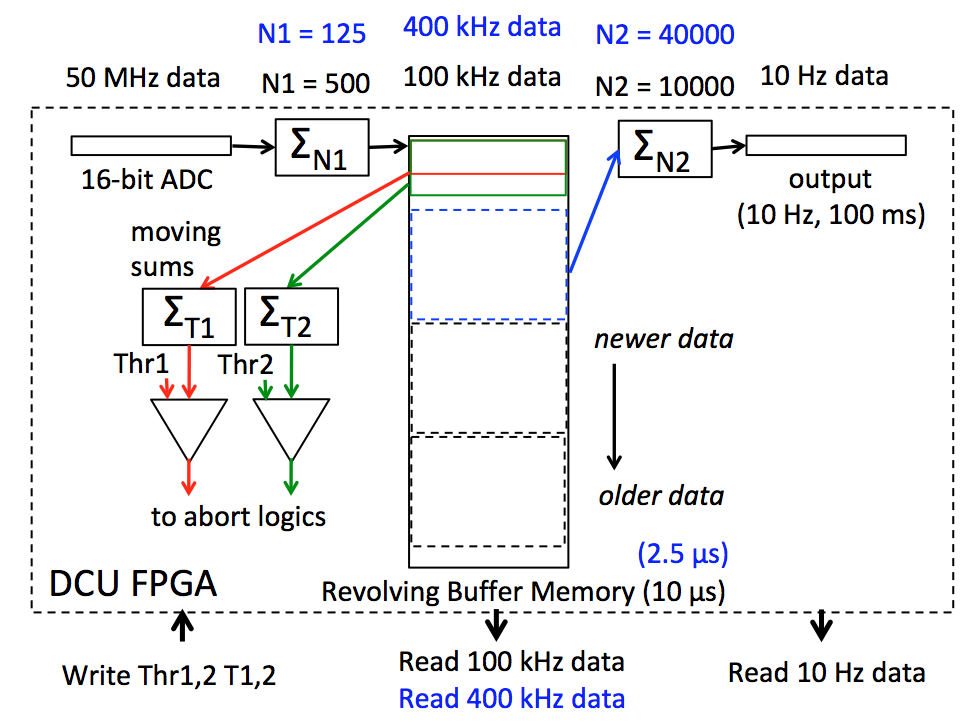}
\caption{Diamond control unit (DCU): simplified block diagram of the FPGA firmware.}
\label{fig:DCU_firmware}
\end{center}
\end{figure}

The diamond currents are amplified by trans-impedance amplifiers, digitised by a 16-bit ADC~\cite{ref_ADC} at $50$~Msamples/s, and processed by FPGAs in the DCUs, as shown in the simplified block diagram of Figure~\ref{fig:DCU_firmware}.

Three amplifier gain values can be selected by resistors in the feedback loop of the front-end operational amplifier~\cite{ref_opamp}, to provide three different current measurement ranges, indexed as 0, 1, and 2 in the following. The analogue bandwidth at the lowest gain (range 2) is about $10$~MHz, matched to large and fast signals; it is reduced to the order of $10$~kHz at the highest gain (range 0) used for monitoring smaller signals at $10$~Hz.

The oversampling 16-bit ADC followed by digital integration performs similarly to slower ADCs with a larger number of bits. At the design stage, this solution was preferred for its flexibility in choosing the level of digital integration.

The sum of 125 samples is obtained every 2.5~$\mu$s, at 400 kHz. These sums, called in the following ``400 kHz data", are written in a 4 Gbit DDR circular buffer memory. These numbers refer to an upgraded version of the firmware, in use since January 2020; in the previous version, sums of $500$ samples were obtained at $100$~kHz, every $10$~$\mu$s.

Two ``moving sums" of ``$400$~kHz data" are updated for each diamond sensor at each memory cycle, by subtracting the oldest added value of ``400 kHz data" and adding the newest one. Corresponding to integrated doses over programmable time intervals, they are compared with programmable thresholds, initially intended to provide abort requests on two different time scales: (1) ``fast" or ``acute" aborts for sudden large spikes in beam losses, typically in the millisecond time range, and (2) ``slow" or ``chronic" aborts for moderate, but increasing beam losses. Operations experience showed that abort requests are delivered mainly on quickly rising beam losses. Integration-time windows were therefore shortened, producing ``very fast" and ``fast" aborts, as described in Section~\ref{sec:beamaborts}.

If a moving sum exceeds the corresponding programmed dose threshold, a logical signal is generated; in total, eight signals are available per DCU. Individual masks can be applied to exclude noisy channels, if needed. ``Abort request" signals are generated separately for LER and HER then sent to SuperKEKB, when a programmed minimum number of unmasked signals above threshold is reached. After activating the beam abort kicker magnets, the accelerator control system broadcasts ``SuperKEKB Abort" signals, which are input to DCUs.

Incoming ``SuperKEKB LER Abort" and ``SuperKEKB HER Abort" signals from SuperKEKB stop the memory writing. The $400$~kHz data can then be read out and written to a file for ``post-abort" analysis of the beam losses preceding the abort. 

Data at $400$~kHz are further added up in groups of 40000, to provide sums of 5000000 ADC values that can be read out at $10$~Hz (``$10$~Hz data"). 

Prototypes were extensively characterized during commissioning: Table~\ref{tab:C2_DCU_ranges} shows the measured characteristics of the three options for current-measurement ranges, selected by changing the amplifier gain, as implemented in the final production version of the DCUs. The range 0 ($36$~nA) allows precise monitoring of relatively small beam losses, while the range 2 ($4.5$~mA) avoids saturation in the detection of large radiation spikes: the latter is used if the DCU is specifically dedicated to provide beam abort requests. The quoted noise is measured in the complete experimental set up during normal accelerator operations. The intrinsic electronics noise, measured in a test bench in a laboratory, is about 10 (4) times better at 100 kHz (10 Hz).

\begin{table}
\centering
\caption{
Options for current-measurement ranges of diamond control units, with the corresponding rms noise values, measured in $100$~kHz data (third column) and in $10$~Hz data (fourth column). 
}
\label{tab:C2_DCU_ranges} 
\begin{tabular}{llll}
\hline\noalign{\smallskip}
Range & Current & Rms noise & Rms noise \\
index  & range  & @100 kHz & @10 Hz \\
\noalign{\smallskip}\hline\noalign{\smallskip}
0 & $36$~nA & $0.23$~nA & $0.8$~pA \\
1 & $9$~$\mu$A & $3$~nA & $70$~pA \\
2 & $4.5$~mA & $0.22$~$\mu$A & $40$~nA \\
\hline\noalign{\smallskip}
\noalign{\smallskip}
\end{tabular}
\end{table}

\subsection{Slow control}
\label{subsec:slowcontrol}
The slow control handles the DCU configuration and the monitoring data in EPICS 3.14~\cite{ref_epics} code. Process variables (PV) are employed to regularly query $10$~Hz data. After subtraction of pedestals, which are ADC values averaged over a prior no-beam time, and multiplication with calibration factors, values of dose-rates are obtained. All dose rate PVs are archived. 

A finite-state machine controls the various phases, with transitions between configurations, $10$~Hz data readout, and buffer memory readout for post-abort analysis. This last state transition is triggered by the arrival of ``SuperKEKB Abort" signals, which stop the advancement of the buffer memory pointers. After completion of buffer memory readout, the system resumes the $10$~Hz data readout.

Time-dependent dose rates are displayed in both the Belle II and the SuperKEKB main control rooms (Figure~\ref{fig:RadView}). Also the pre-abort history of dose rates with 2.5~$\mu$s time resolution is available to operators for post-abort analysis of the beam conditions, as discussed in Section~\ref{sec:beamaborts}. Individual diamond detectors are labelled as a\_b\_c according to their position; \textup{a = BP, SVD, QCS} describes the location on beam pipe (BP), SVD support cones, or QCS bellows; \textup{b = FW, BW} denotes the forward or backward position with respect to the interaction point, in the electron beam direction; c represents the azimuthal angle in degrees, in the standard Belle II laboratory reference frame.

\section{Dose rate measurements}
\label{sec:doserates}
The dose rates from the 28 diamond detectors, measured and archived at 10 Hz, provide on-line monitoring of beam losses and accelerator-related backgrounds, and are correlated with the data of Belle II inner detectors. We describe here the main features and results of these measurements.

\subsection{Pedestals and noise}
\label{subsec:pedestals_noise}
\begin{figure*}[htbp]
	\begin{center}
	    \includegraphics[width=6.5cm]{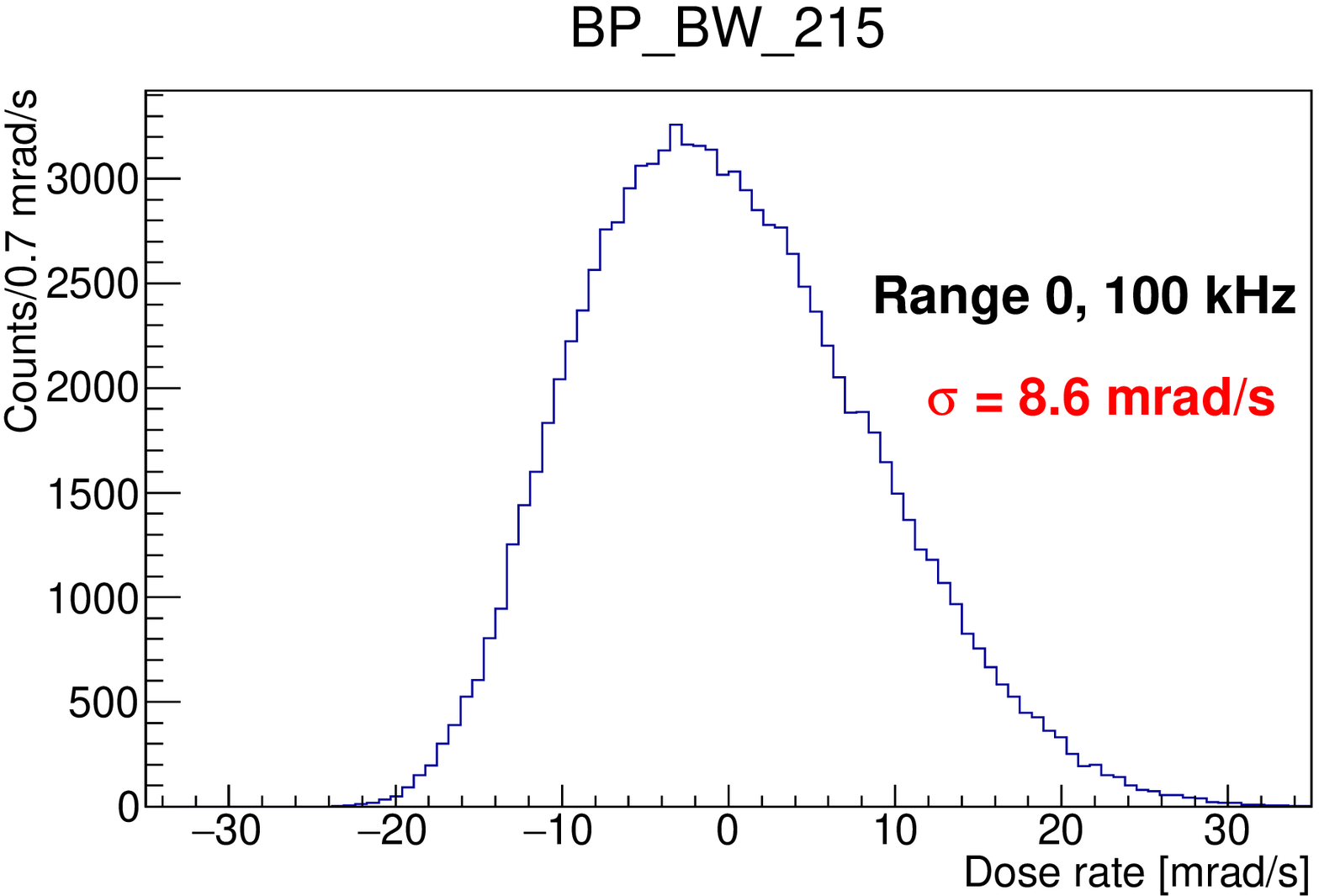}
		\includegraphics[width=6.5cm]{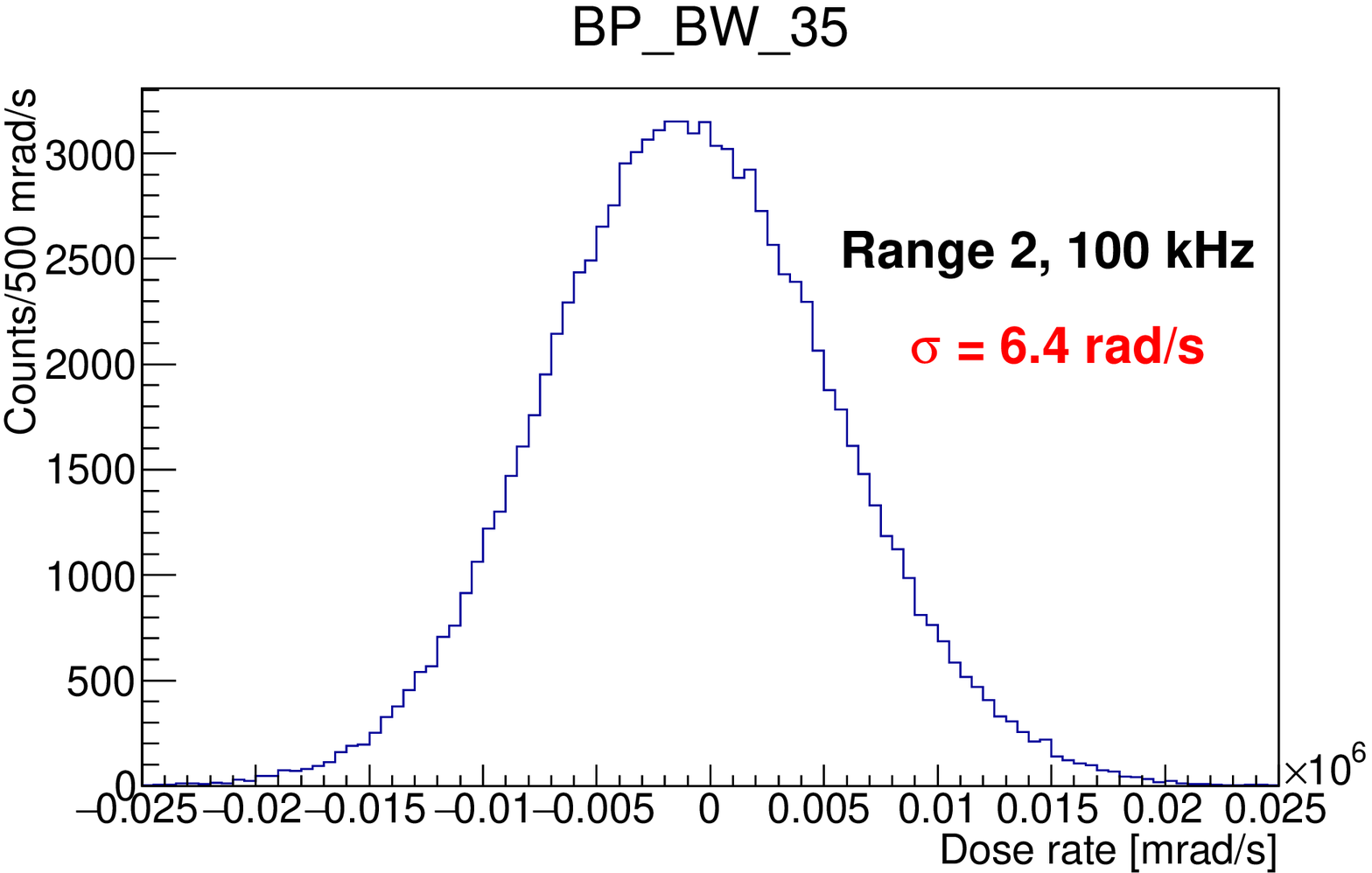} \\
		\includegraphics[width=6.5cm]{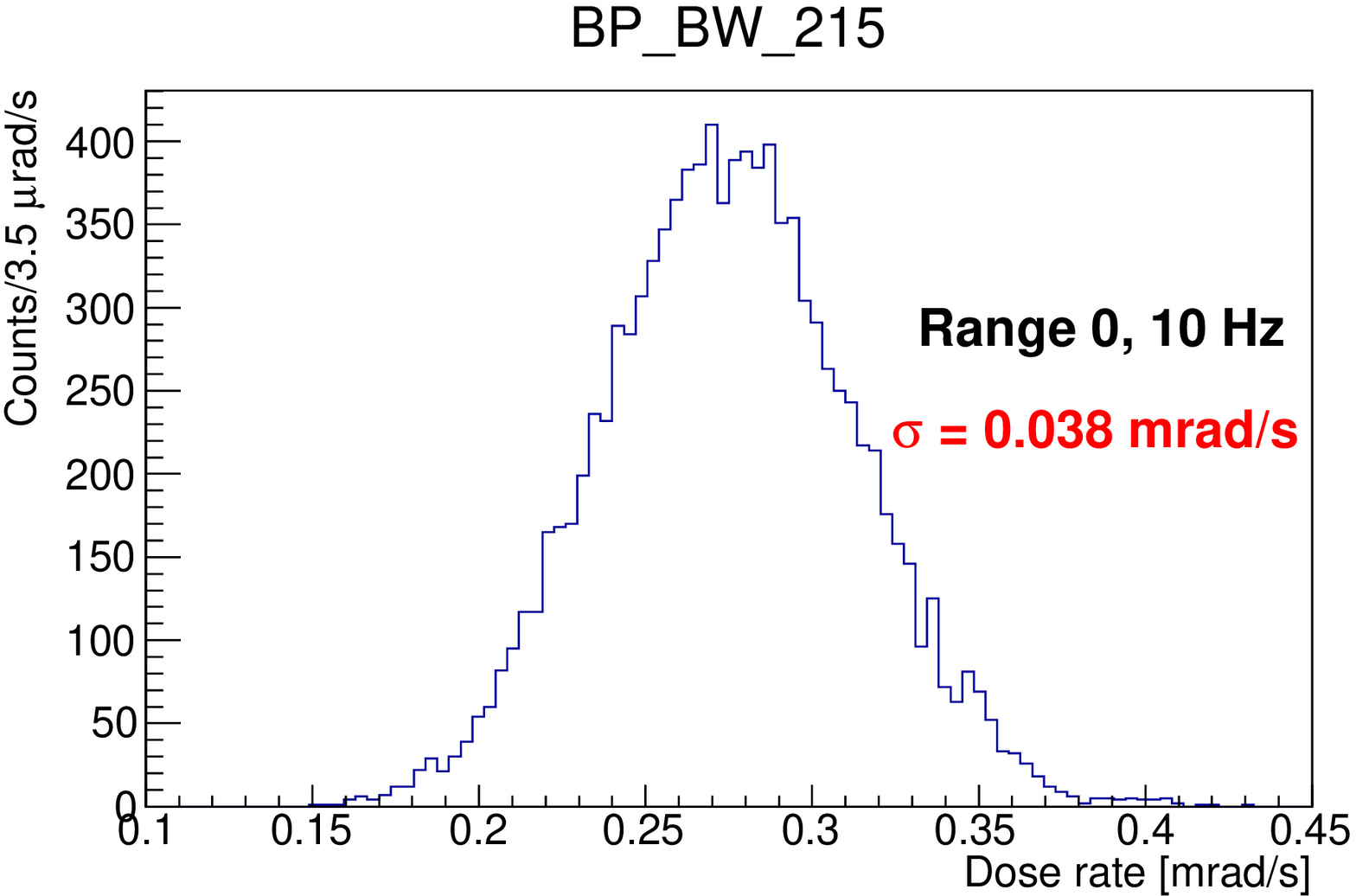}
		\includegraphics[width=6.5cm]{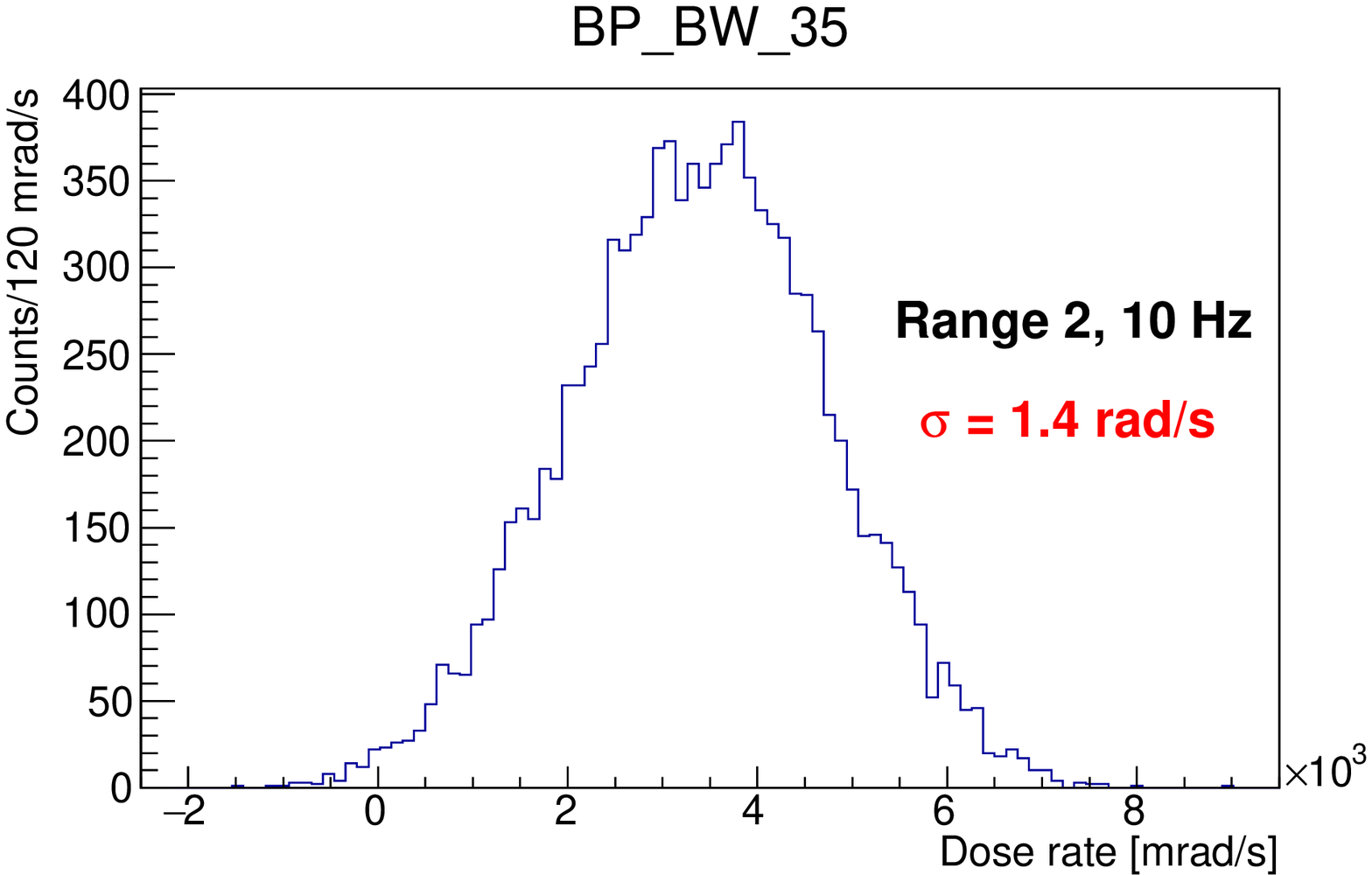}
		\caption{Examples of noise measurements performed without circulating beams for diamond detectors installed on beam pipe, expressed in mrad/s: (top left) range 0, $100$~kHz; (top right) range 2, $100$~kHz; (bottom left) range 0, 10 Hz; (bottom right) range 2, 10 Hz. Entries in the top plots are read from the $100$~kHz buffer memory, spanning $1$~s (100000 entries), pedestal-subtracted and converted from ADC to dose-rate units; the slight asymmetry in these distributions is due to an intrinsic cutoff of the largest negative fluctuations in the almost-unipolar front-end range. Entries in the bottom plots are 10 Hz data collected in 17 minutes ($\sim$10000 entries); the shift of the average from zero corresponds to a pedestal drift since the previous measurement.}
		\label{fig:noise}
	\end{center}
\end{figure*}

Offsets (pedestals) and noise of each readout channel are obtained by repeated measurements without circulating beams. A few minutes are sufficient to precisely determine the central value and noise value for the chosen current-measurement range, from histograms similar to the examples in Figure~\ref{fig:noise}. Table~\ref{tab:noise} gives a summary of measured noise for a typical diamond sensor calibration factor $k=30$~(mrad/s)/nA.

\begin{figure}[htbp]
	\begin{center}
		\includegraphics[width=8cm]{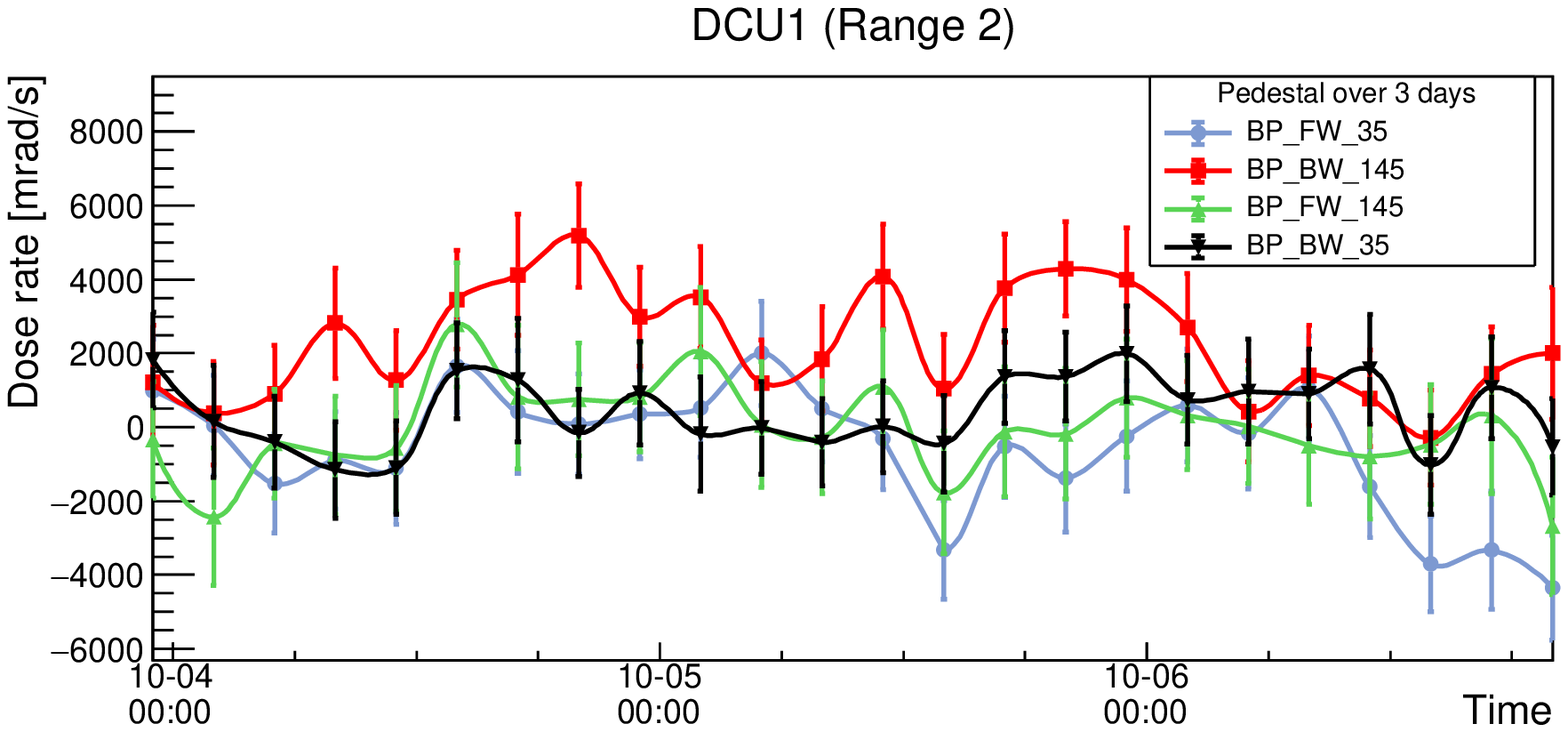}
		\includegraphics[width=8cm]{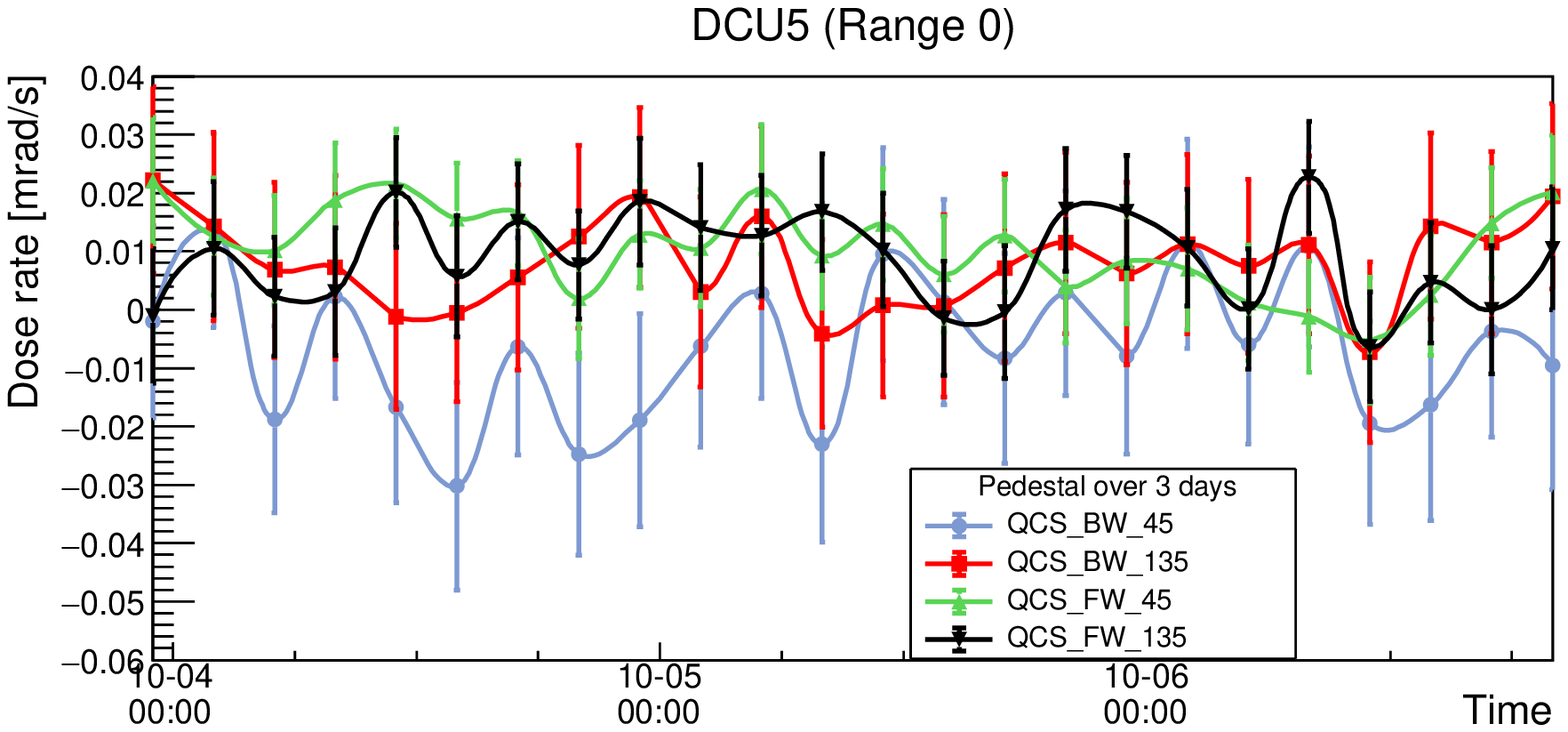}
		\caption{Example of time evolution of pedestals and noise (mrad/s) over three days without beams. The central values represent pedestals (averages of 10 Hz data) evaluated in short time intervals of 10 minutes every three hours; error bars show the standard deviations of the corresponding approximately Gaussian distributions, representative of noise, for measurements performed in range 2 (top) and in range 0 (bottom).}
		\label{fig:NoisePed}
	\end{center}
\end{figure}

\begin{table}
\centering
\caption{
Typical measured rms-noise values expressed in dose rate units for the three measurement ranges; the approximate noise for $100$~kHz data in range 2 is relevant for the ``very fast abort" threshold with $10$~$\mu$s integration time (Section~\ref{subsec:abort_thresh}).
}
\label{tab:noise} 
\begin{tabular}{llll}
\hline\noalign{\smallskip}
Range   & Dose rate           & Rms noise           & Rms noise        \\
index   &  range               & @100 kHz           & @ 10 Hz          \\
\noalign{\smallskip}\hline\noalign{\smallskip}
0      & $1.1$~rad/s     & $6.8$~mrad/s     & $0.024$~mrad/s      \\
1      & $270$~rad/s     & $90$~mrad/s      & $2.2$~mrad/s         \\
2      & $140$~krad/s    & $7$~rad/s        & $1.2$~rad/s         \\
\hline\noalign{\smallskip}
\noalign{\smallskip}
\end{tabular}
\end{table}

The measured noise includes a random component that scales with the data averages from 100 kHz to 10 Hz, and common-mode components, picked up from the environment, which are only partially averaged out. Common-mode noise depends on the selected range and shows a pattern correlated with the detector cabling: substantial attenuation is obtained by placing ferrite bead filters on the DCU input signal cables.

The measured average values (pedestals) are mostly stable (Figure~\ref{fig:NoisePed}), drifting typically less than one or two times the standard deviations of beam-off pedestals during the time interval between programmed accelerator beam-off periods: a weekly or bi-weekly update of pedestals is sufficient for most purposes, including the pedestal subtraction for monitoring of instantaneous dose rates. For off-line computations requiring a more accurate subtraction, we use the data archived during short accidental beam-off periods to determine smaller pedestal shifts more frequently.

\subsection{Instantaneous dose rates}
\label{subsec:inst_doserates}
Diamond current signals, converted to dose rate units after pedestal subtraction, are continuously monitored in all accelerator conditions. They are particularly relevant as indicators of beam quality during injection and collimator tuning, and for beam optics studies. %Figure~\ref{fig:DoseRates} shows the sensitivity to very small variations in beam losses.

\subsection{Integrated dose}
\label{subsec:integrated_dose}
\begin{figure}[htbp]
	\centering
    	\subfloat[Beam pipe]{{\includegraphics[width=7.5cm]{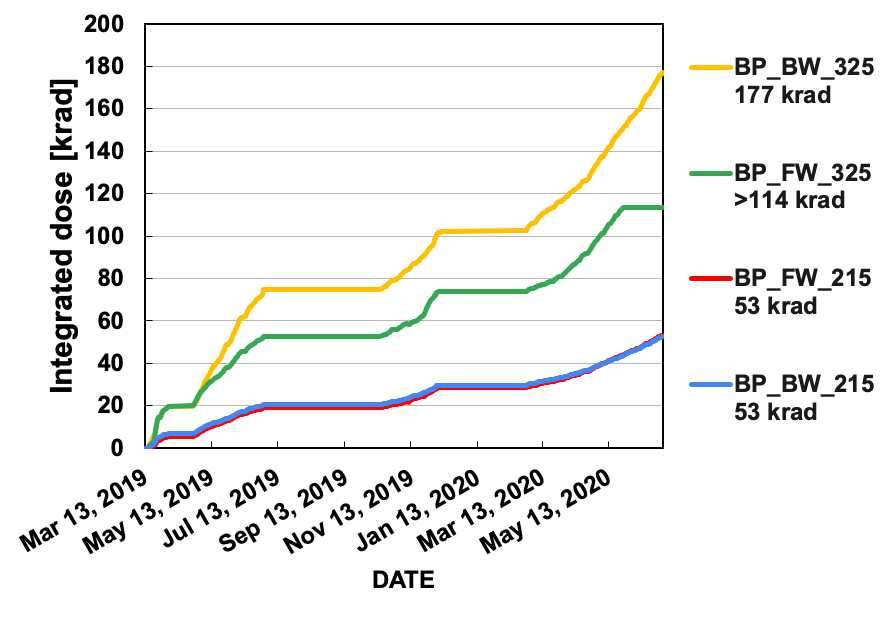} }}%
    	\hfill
		\subfloat[QCS backward]{{\includegraphics[width=7.5cm]{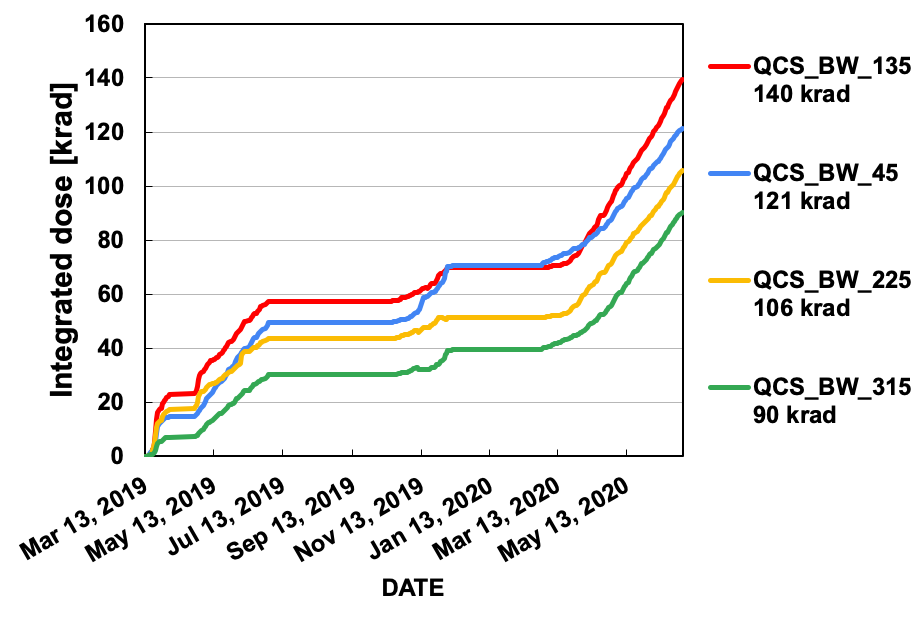} }}%
		\hfill
		\subfloat[QCS forward]{{\includegraphics[width=7.5cm]{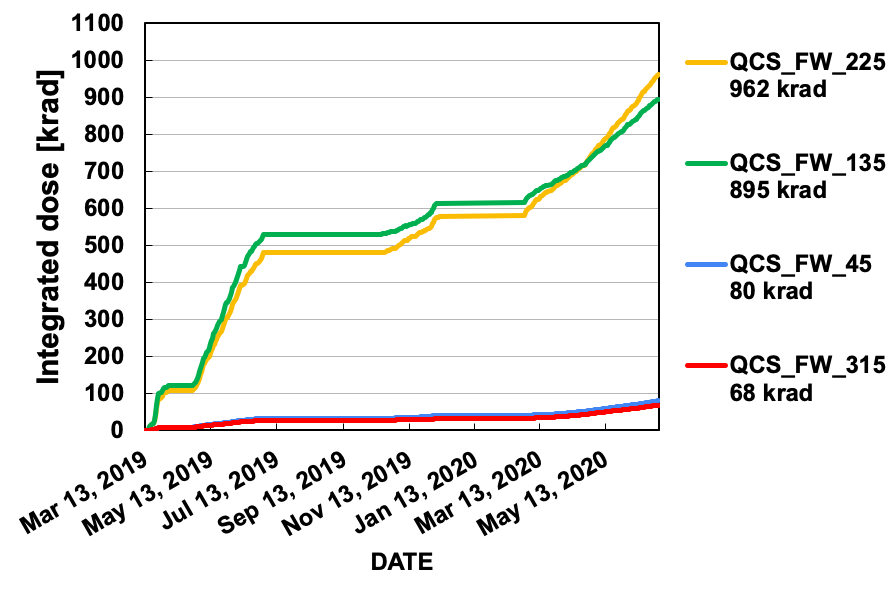} }}%
	\caption{Integrated dose during Phase 3 (March 13, 2019 to July 1, 2020), in diamond detectors located on the beam pipe (a) and close to the final-focusing superconducting magnets in the backward (b) and forward (c) regions. 
	}
	\label{fig:IntDose}
\end{figure}

The integrated radiation doses are approximately computed on-line, using the $10$~Hz monitoring data, and made available on a daily basis. Small pedestal shifts are then determined (Section~\ref{subsec:pedestals_noise}) and used off-line for a more precise estimate of the integrated dose, as reported in Figure~\ref{fig:IntDose} for the most exposed diamond detectors, located on the beam pipe (BP) and the QCS bellows. The diamond detectors mounted on SVD cones received radiation doses lower by almost an order of magnitude, concentrated mostly in the horizontal plane of the circulating beams. 

The most exposed diamond detector QCS\_FW\_225 received an estimated total dose of about $960$~krad in Phase 3 until July 2020. 

The pre-amplifier of diamond detector BP\_FW\_325 was damaged during a beam incident with very large losses in May 2020, as shown by the dose for that channel in Figure~\ref{fig:IntDose}(a), no longer increasing after the incident. The pre-amplifier was repaired and the diamond read-out channel recovered for the next run period.

\subsection{Accelerator tuning and beam backgrounds studies}
\label{subsec:backgrounds_studies}
During Phase 2 several studies were dedicated to investigate the origin of beam-related backgrounds, by correlating detector outputs with the accelerator conditions. Diamond detectors contributed to these studies, as shown in the following examples~\cite{Bassi:2018mdf}.

\begin{figure*}[htbp]
	\centering
    	\includegraphics[width=12cm]{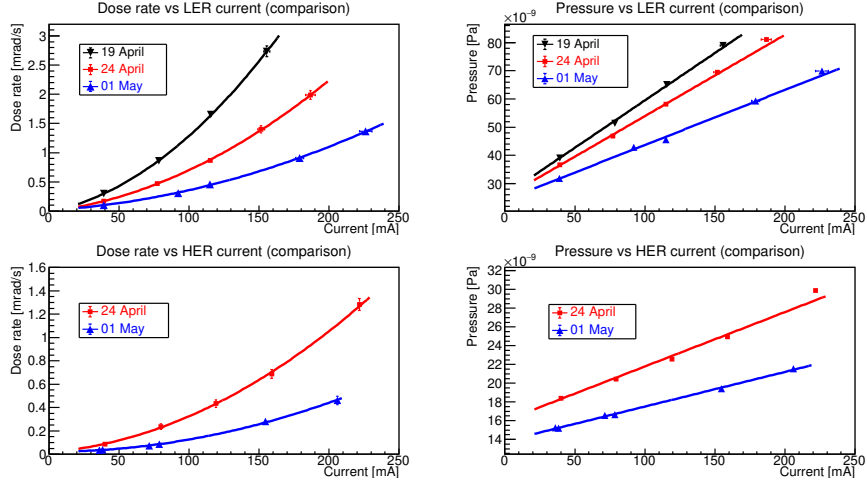}
	\caption{(left) Dose rate of the diamond detector with highest signal as a function of beam currents, separately for LER and HER, during single-beam measurements at different dates in Phase 2; (right) residual gas pressure increase with the beam current, during the same measurements.}
	\label{fig:Dose_vs_BeamCurrent}
\end{figure*}

The left panels in Figure~\ref{fig:Dose_vs_BeamCurrent} show the dose rates, measured by the diamond detector on the beam pipe showing the highest signal, as functions of the circulating beam current, separately for LER and HER, at different dates. 

Single beam backgrounds, shown here, are dominated by two types of contributions. Beam losses due to beam interactions with the residual gas, the so called beam-gas backgrounds, scale with the product of the beam current and the residual gas pressure. Touschek background, due to intra-bunch scattering, scales quadratically with the beam current.

Two features are evident in these plots: a progressive lowering of dose rates with increasing date and a quadratic dependence of dose rate on beam current.

The former feature is explained by several factors. In particular, the accelerator vacuum was improved by ``beam scrubbing" the vacuum pipes, with consequent decrease of the beam losses due to beam interactions with the residual gas, and collimator tuning contributed to limit the beam losses. 

The quadratic dependence from beam current, valid for Touschek losses, is qualitatively explained also for the beam-gas background, since the residual gas pressure is increasing with the beam current, as shown in the right panels of Figure~\ref{fig:Dose_vs_BeamCurrent}.

\begin{figure}[htbp]
	\centering
    	\includegraphics[width=7cm]{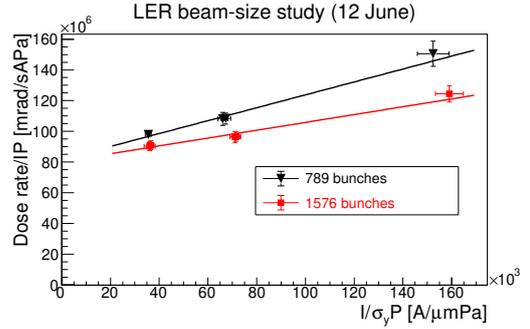}
	\caption{Dose rate of one diamond detector as a function of the reciprocal of the LER beam size. Both variables are normalized using the constant LER beam current and the effective vacuum pressure. The linear fit extracts the beam-gas (constant) and Touschek (linearly increasing) contributions, as shown in equation~\ref{eq:Dose_vs_BeamSize}. Sharing the same beam current in a larger number of bunches (1576 instead of 789) decreases the Touschek contribution, as expected.}
	\label{fig:Dose_vs_BeamSize_1}
\end{figure}

\begin{figure}[htbp]
	\centering
    	\includegraphics[width=7cm]{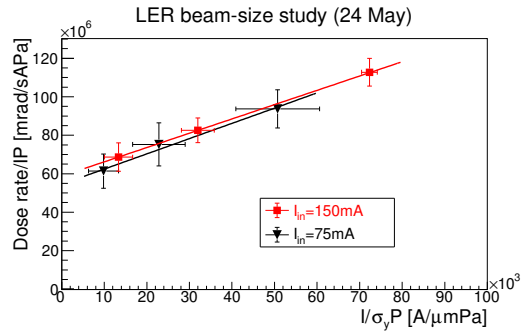}
	\caption{Dose rate in one diamond detector as a function of the reciprocal of the LER beam size, similar to Figure~\ref{fig:Dose_vs_BeamSize_1}, for two different LER beam currents. The two sets of measurements overlap, as expected for the dose rate normalized on the beam current.}
	\label{fig:Dose_vs_BeamSize_2}
\end{figure}

A quantitative analysis separates the contributions of beam-gas and Touschek backgrounds, using data from dedicated studies with single beams, as functions of beam size and number of bunches at fixed beam current $I$~\cite{Lewis:2018ayu}. Beam-gas backgrounds remain constant in these conditions, while Touschek intra-bunch scattering is expected to increase when the density of particles in each bunch increases, by decreasing the beam size or lowering the number of bunches at the same total current. This is qualitatively shown in Figure~\ref{fig:Dose_vs_BeamSize_1} and Figure~\ref{fig:Dose_vs_BeamSize_2} and quantitatively expressed in the parameterization~~\cite{Lewis:2018ayu}
\begin{equation}
\frac{dD}{dt} \frac{1}{P_{e} I} = S_{bg} Z_{e}  + S_{T} \frac{I}{P_{e} \sigma_{y}} ,
\label{eq:Dose_vs_BeamSize}
\end{equation}
where $P_{e}$ is the residual gas average effective pressure, $Z_{e}$ the effective atomic number of the gas mixture, $\sigma_{y}$ the vertical beam size, $S_{bg}$ and $S_{T}$ coefficients related to the relative contributions of beam-gas and Touschek backgrounds respectively.

%Figure~\ref{fig:BeamLifetime_vs_BeamSize} shows...
%
%\begin{figure}[htbp]
%	\centering
%   	\includegraphics[width=7.5cm]{figures/fig_BeamLifetime_vs_BeamSize_Ph2.png}
%	\caption{Beam lifetime vs beam size (Phase 2).}
%	\label{fig:BeamLifetime_vs_BeamSize}
%\end{figure}

Similar studies have been repeated in Phase 3 at each major step of the evolution in accelerator conditions, such as increasing beam currents, reducing beam size at the interaction point, optimization of collimator settings and of the parameters of continuous injection.

Dedicated studies, based on the combination of the pattern of signals from diamond detectors and the extrapolation of particle tracks from Belle II tracking detectors, were used to identify specific background sources such as those due to beam tails ``scraping" the vacuum-chamber walls. 

``Beam-dust events", characterized by localized pressure bursts accompanied by radiation bursts, often triggered beam aborts and losses of operation time, more frequently in the initial operation of the accelerator. They were studied by looking for large sudden signals from diamonds, in coincidence with other detectors, and interpreted as collisions between beams and small particles such as dust, coming off the vacuum-chamber material.

\subsection{Comparison with simulations}
\label{sub:simulations}
Simulations of beam-related backgrounds are needed for two main purposes. The space and time distribution of the radiation field in the interaction region correlates the radiation doses measured by diamond sensors with the radiation dose received by the neighboring Belle II inner detectors, PXD and SVD. Moreover, comparisons between simulations and measurements validate the simulations, which can be then used to identify and estimate the background sources and to extrapolate their effects in the future accelerator conditions.

Single-beam background simulation studies start with the simulation of a single circulating beam and its interactions (Touschek, Bremsstrahlung, and Coulomb scattering) based on the SAD code~\cite{ref_SAD}, followed by tracking of lost particles and secondaries in the beam pipe and the surrounding material with the GEANT4 simulation of Belle II~\cite{Agostinelli:2002hh}. The geometry and materials of the diamond detectors are included in the simulation. The energy deposited per unit time by particles crossing each diamond detector is converted into a dose rate, using the measured calibration constants.

\begin{figure}[htbp]
	\centering
	\includegraphics[width=6.5cm]{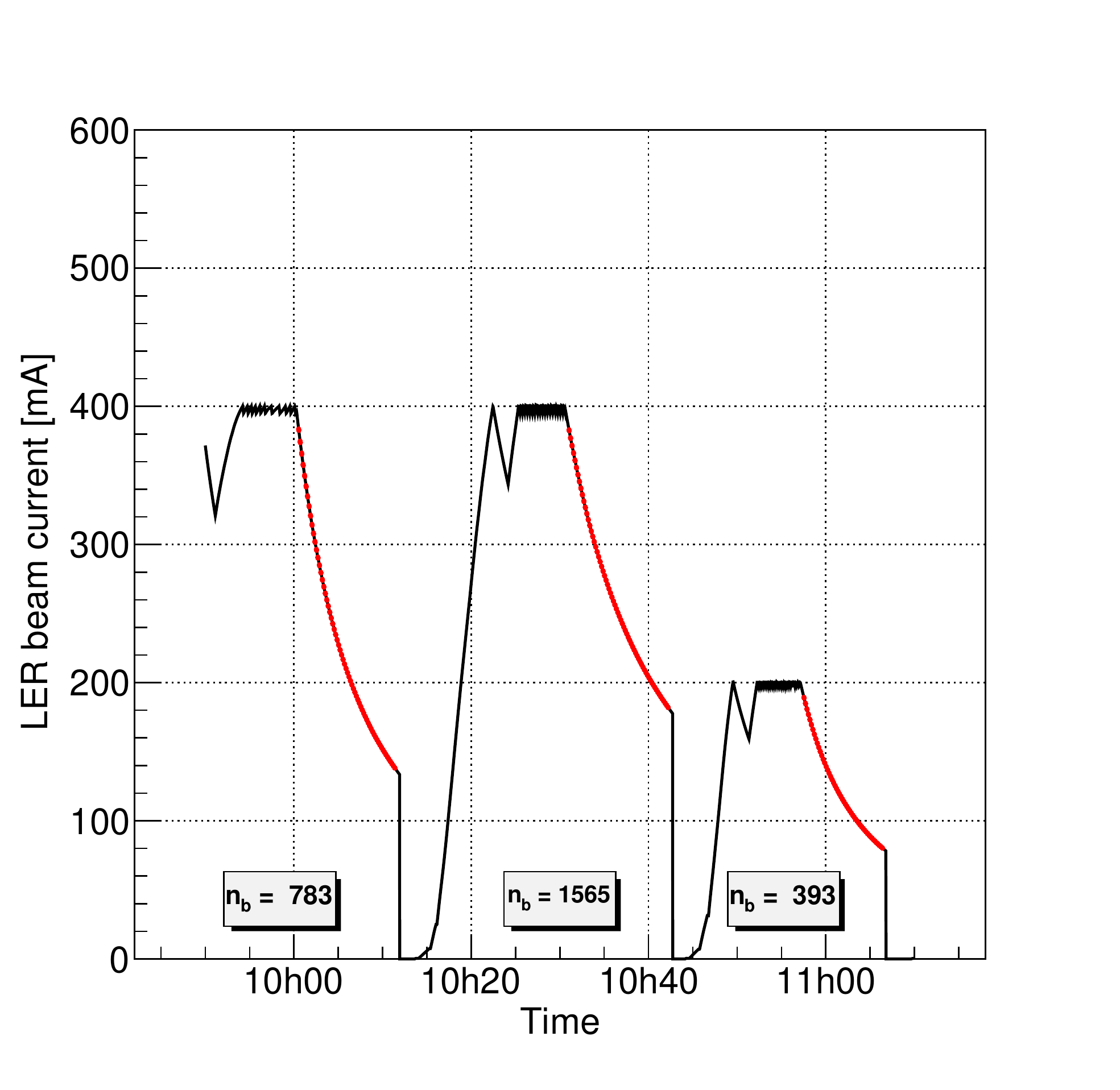} \\
	\includegraphics[width=6.5cm]{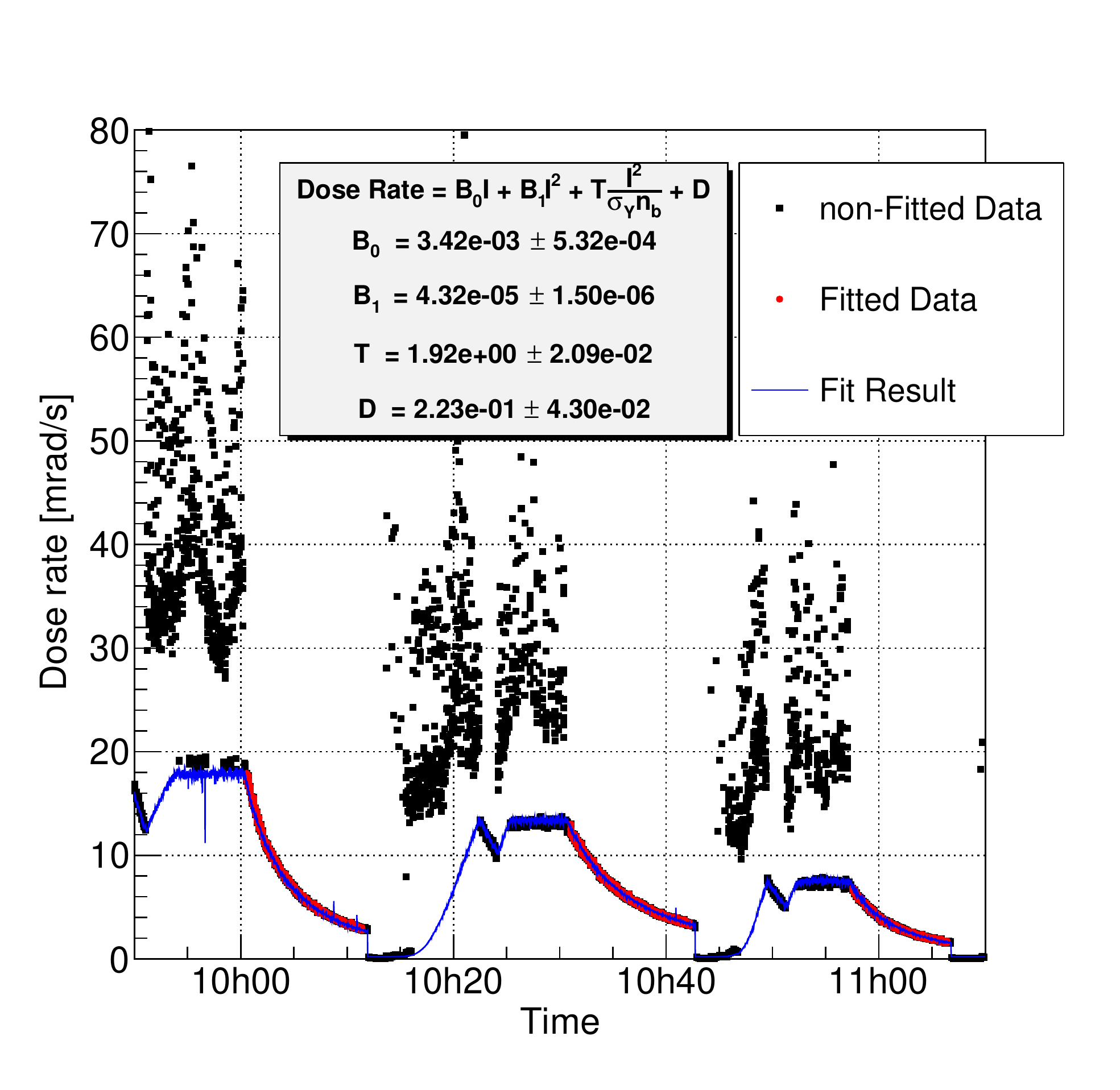} \\
	\includegraphics[width=6.5cm]{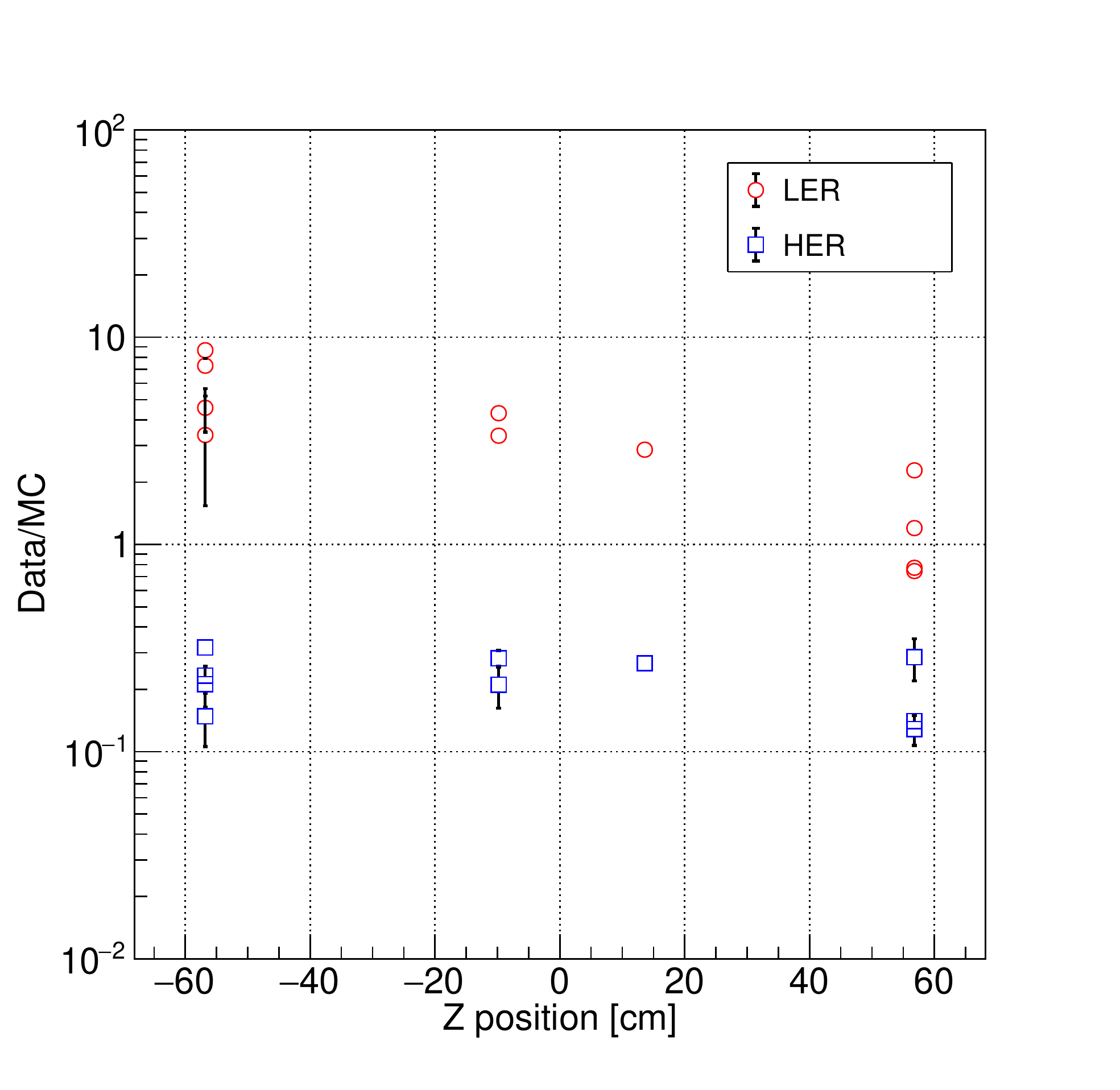}
	\caption{Comparison between dose rate data and simulations: an example from a study of backgrounds from single LER beam. (Top) LER beam current $I$ as a function of time; red line: beam-current decay period; $n_{b}$: number of bunches. (Middle) Corresponding dose rate from beam losses recorded by QCS\_FW\_135; black squares: non-fitted data, red circles: fitted data; blue line: fit results for the study. (Bottom) Ratio of recorded and simulated data for eight QCS and three BP diamond detectors, as a function of their positions; red circles: LER; blue squares: HER.
	}
	\label{fig:simulations}
\end{figure}

Simulations are repeated for the specific beam optics and collimator settings of successive data taking periods and accelerator tuning studies. Figure~\ref{fig:simulations} shows an example from LER single-beam background studies in June 2020. After initial injection, the beam is kept for some minutes at constant current with continuous injection, and then the current decays without injection (red line in Figure~\ref{fig:simulations}(top)). This pattern is repeated three times, with different number of bunches $n_{b}$ filled into the machine (Figure~\ref{fig:simulations}(top)). The beam losses recorded by the diamond detectors follow this pattern closely; the injection-related background is well separated and can be excluded from a fit (blue line in Figure~\ref{fig:simulations}(middle)) taking into account beam current $I$, number of bunches, vertical beam size $\sigma_{Y}$, beam-gas ($B_{0}$ and $B_{1}$) and Touschek $T$ background components, as shown in the insert of the figure. The data agree with simulations within an order of magnitude (Figure~\ref{fig:simulations}(bottom)); the comparison helps understanding the effect of collimators, alignments and other accelerator parameters in the simulation.

\subsection{Saturation effects}
\label{subsec:saturation}
In the continuous injection mode of SuperKEKB, the intensities of the circulating beam bunches are topped up repeatedly over short time intervals, at a frequency of up to $25$~Hz. Immediately after each injection, beam losses in the interaction region grow until the beam oscillations due to the injection perturbation are dumped away. As a result, beam losses and diamond detector signals vary in time; a significant part of the radiation dose is delivered during short time intervals, adding up to a few milliseconds per second, depending on the specific injection pattern and frequency.

The dose-rate monitor at 10 Hz and the total dose computation are based on digital sums of the 50 MHz ADC data over $100$~ms time intervals (Section~\ref{subsec:electronics}, Section~\ref{subsec:slowcontrol}). If saturation of the current signals occurs in range 0 over much shorter time intervals, it becomes partially hidden by averaging in $10$~Hz data, leading to an underestimate of radiation dose rates and integrated doses.

\begin{figure}[htbp]
	\centering
	    \includegraphics[width=7.5cm]{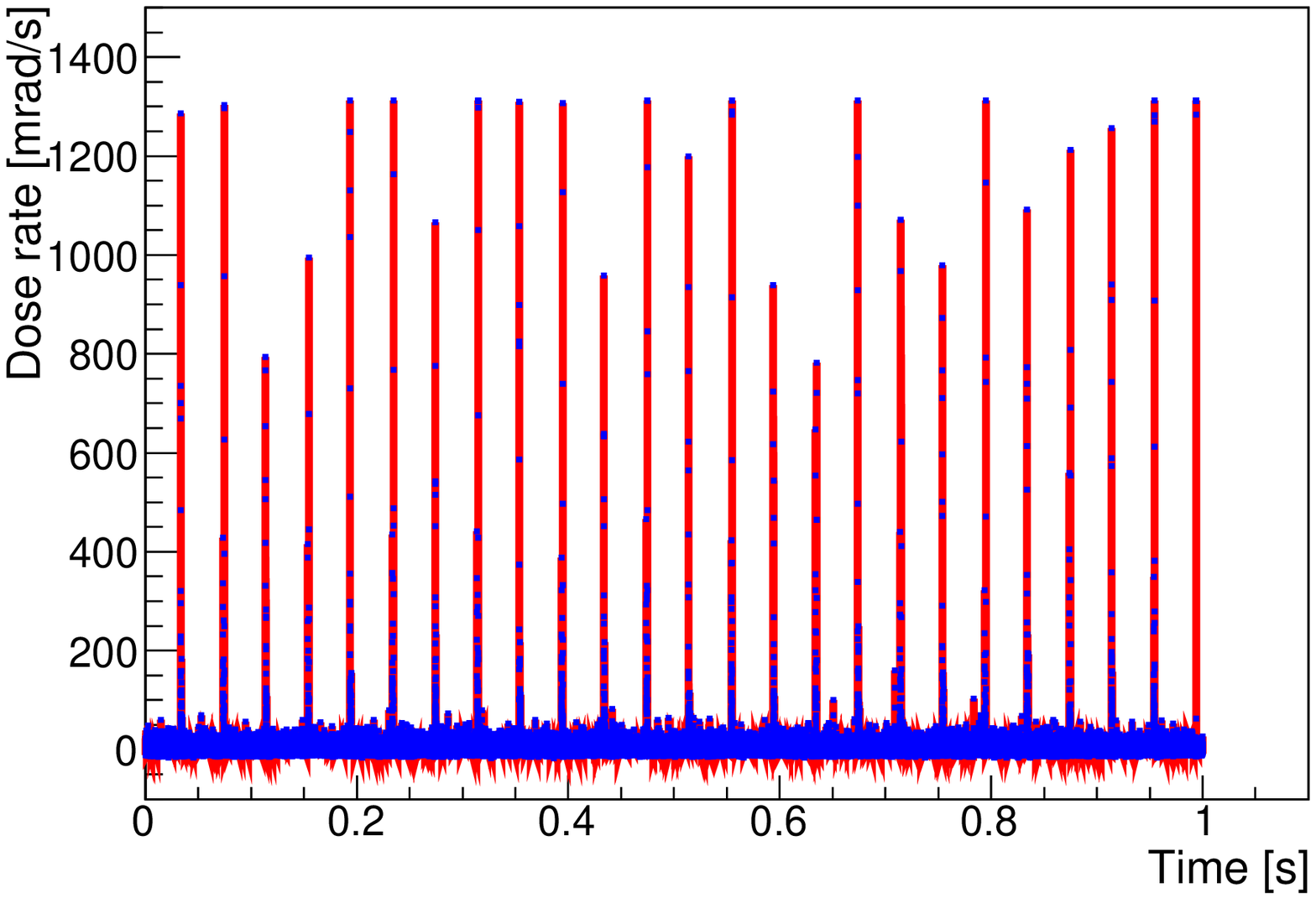}
	    \includegraphics[width=7.5cm]{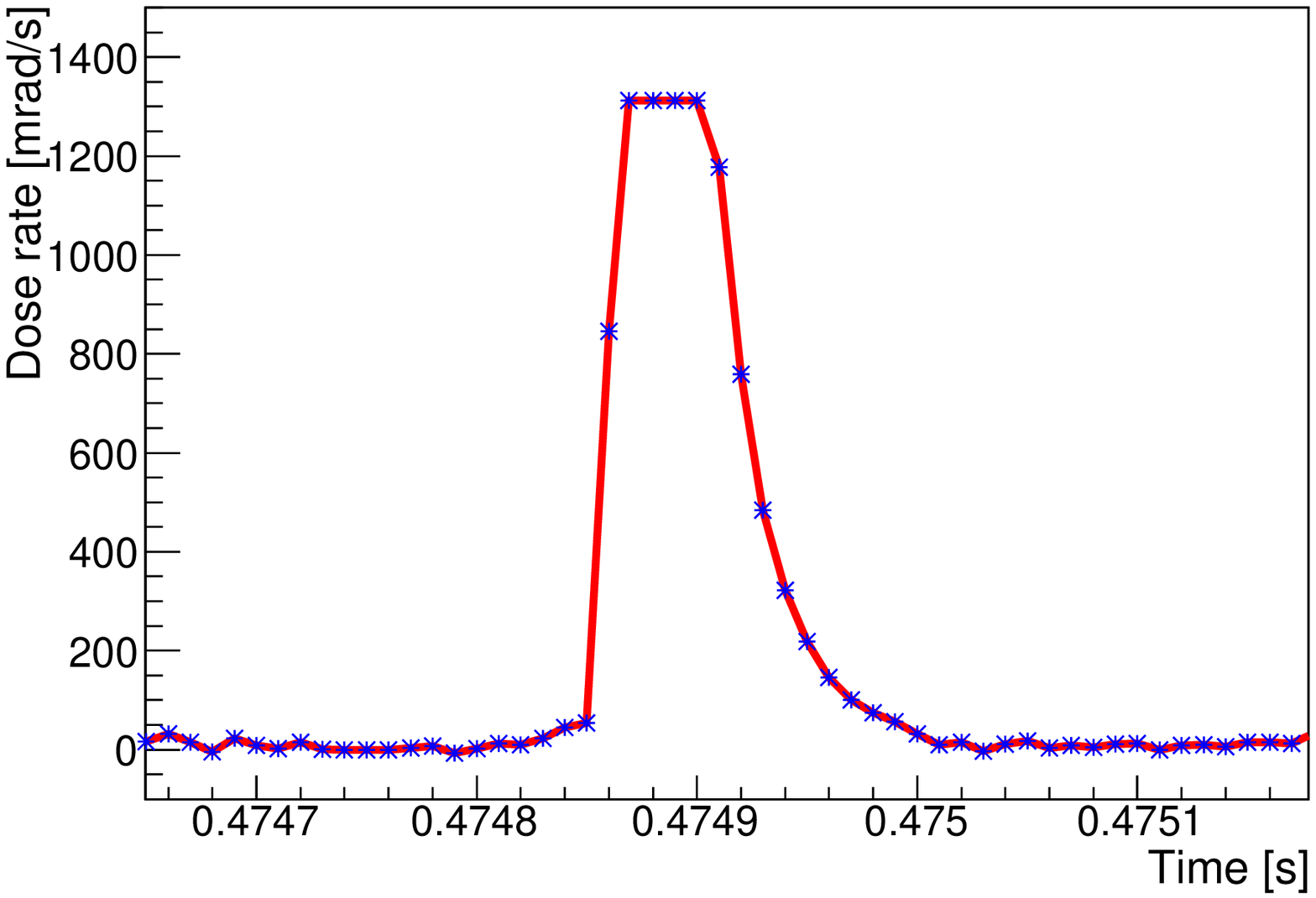}
	\caption{(Top) An example of saturation in range 0, observed in a one-second time interval. The $100$~kHz data are read during continuous injection at $25$~Hz for diamond detector BP\_BW\_325. Ten out of $25$ injection-related peaks have a saturated value of about $1.3$~rad/s, at least for one $10$~$\mu$s sampling each;  (bottom) one of the above peaks saturates the measurement range during $40$~$\mu$s.}
	\label{fig:saturation}
\end{figure}

These effects were studied in November 2019 by repeatedly dumping the DCU buffer memories during continuous injection, to observe the beam loss patterns in 100 kHz data with $10$~$\mu$s resolution, for both range 0 ($36$~nA) and range 1 ($9$~$\mu$A). The example in Figure~\ref{fig:saturation} shows that saturation of range 0 could be observed in some BP and QCS diamond detectors, exposed to higher beam losses. Saturation typically lasted for one or a few 100 kHz samples (10-100~$\mu$s). This observation suggests to switch to range 1 for some DCUs in future operations, to avoid saturation at higher beam currents; an upgrade of the electronics toward expanding the dynamic range is also considered in the long term.

\subsection{SVD occupancy and injection inhibit}
\label{subsec:injection_inhibit}
The occupancy of the inner layer of the SVD double-sided microstrip detector of Belle II is a critical parameter for track reconstruction. The performance of the pattern recognition algorithms deteriorate significantly beyond an occupancy of about $3$~\%. 

\begin{figure}[htbp]
	\begin{center}
	    \includegraphics[width=7.5cm]{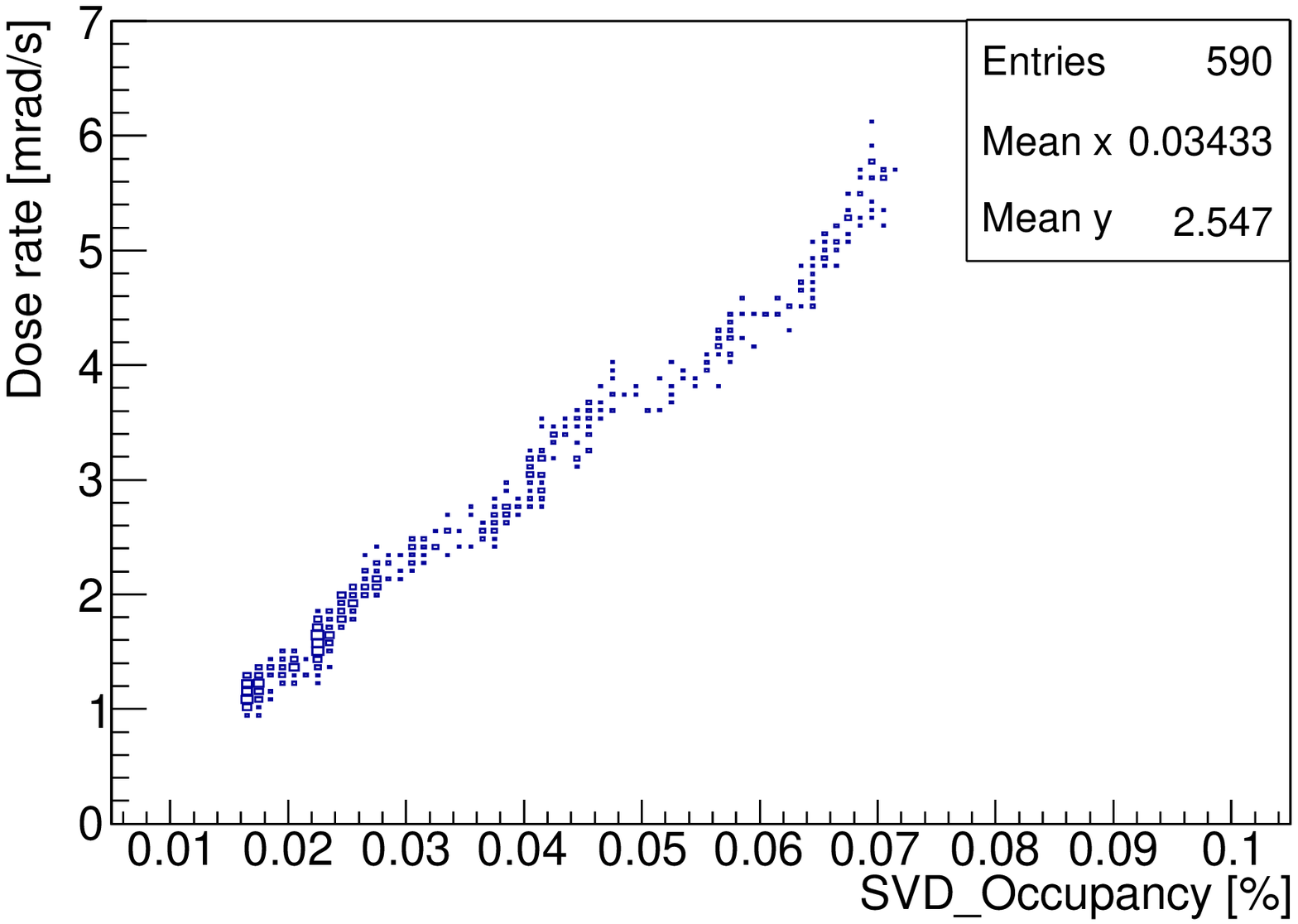}
	    \hfill
	    \vspace{1em}
		\includegraphics[width=7.5cm]{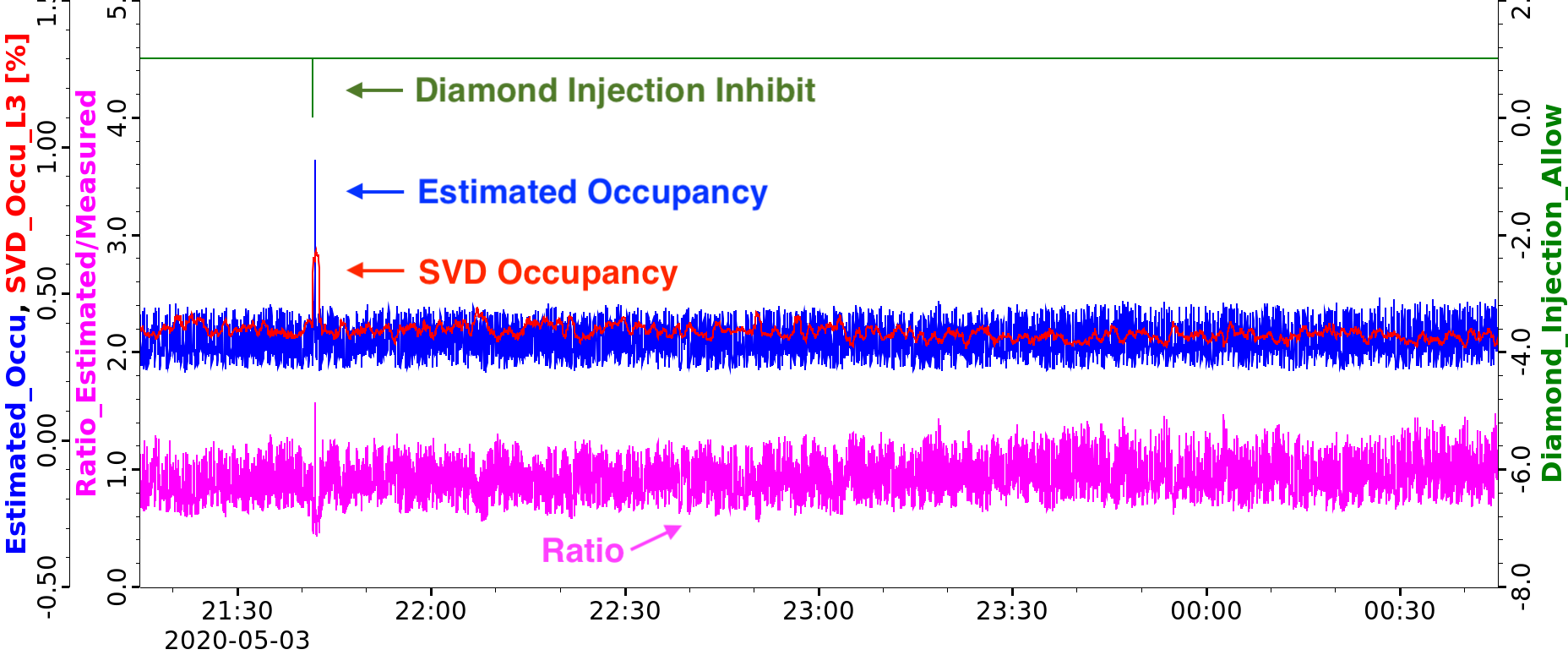}
		\caption{(Top) Correlation between the dose rate in diamond detector BP\_BW\_215 and the occupancy of the inner SVD microstrip detector layer, measured during single-beam background studies with changing beam currents. (Bottom) the beam-pipe diamond detectors' dose rates measured at 10 Hz give an estimated occupancy (blue), compared with the SVD occupancy (red), measured at 1 Hz. Their ratio (pink) is close to unity and stable in time. On the 3rd of May 2020, the estimated occupancy exceeded the limit and triggered a diamond ``injection inhibit" (green). 
		}
		\label{fig:L3_Occu}
	\end{center}
\end{figure}

An approximately linear correlation was observed between SVD occupancy and $10$~Hz data from BP diamond detectors: an example is shown in Figure~\ref{fig:L3_Occu}. The expected occupancy could therefore be predicted in real time by a combination of BP diamond signals. An ``injection inhibit" flag was prepared and implemented in EPICS, to suspend the continuous injection in large background conditions corresponding to unacceptable occupancy.

\section{Beam aborts}
\label{sec:beamaborts}

$24$ diamond detectors were dedicated to monitoring, and the corresponding DCUs preset on the most sensitive current measurement range 0 (section~\ref{subsec:electronics}). Four detectors mounted on the beam pipe were dedicated for the abort function, with range 2 selected in the corresponding DCU.

\subsection{Abort timing}
\label{subsec:abort_timing}

\begin{figure}[htbp]
	\begin{center}
		\includegraphics[width=7cm]{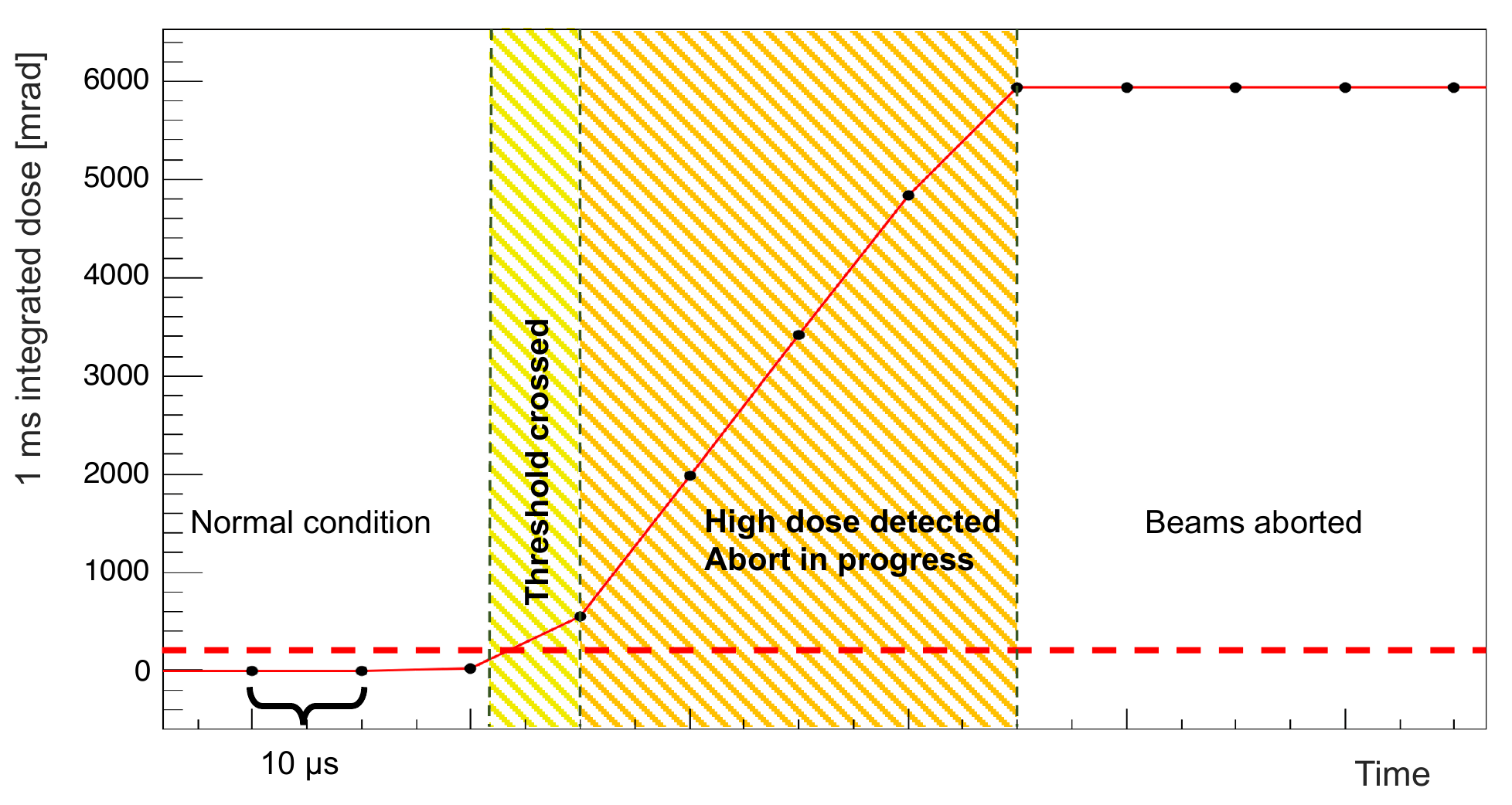}
		\caption{An example of large beam loss, which occurred before the SuperKEKB reduction of abort delays and the DCU firmware improvement described in the text. The integrated dose in a moving window of $1$~ms, updated every $10$~$\mu$s, is shown as a function of time: a dose of $6$~rad is integrated in the almost $50$~$\mu$s interim between abort threshold crossing and beam dump completion.}
		\label{fig:abort_timing}
	\end{center}
\end{figure}

Minimizing the time delays in the abort system is essential for the protection of Belle II and of accelerator components in the interaction region: during the interim between the initial detection of abnormal beam losses and the complete dumping of the beams, radiation levels may increase up to damaging values~\cite{Manfredi:2019lna}, as shown in the example of Figure~\ref{fig:abort_timing}. 

On the accelerator end, optical fibres are used for long-distance propagation of abort signals. Their propagation delay was minimized by optimizing the signal paths in the accelerator central control system; in 2020, a second ``abort gap"~\cite{Mimashi:2014dya} was also introduced in the circulating bunch trains, to effectively reduce the waiting time for the activation of abort kicker magnets.

The generation of abort-request signals in the diamond-based system was initially cycling every $10$~$\mu$s, to compare the moving sums of data stored at $100$~kHz with the abort thresholds. At the beginning of 2020, the modified DCU firmware (section~\ref{subsec:electronics}) contributed to the delay-minimization effort, by storing data in the buffer memory at $400$~kHz and performing the abort-threshold comparison every $2.5$~$\mu$s. As a result, the internal delay of the DCU between a large step-like fast-rising input signal and the delivery of the output abort request is halved to about $6$~$\mu$s. This time includes both the effect of the analog front-end bandwidth and the overhead of an additional confirmation cycle after threshold-crossing.

\subsection{Abort thresholds settings}
\label{subsec:abort_thresh}

The diamond abort thresholds were adjusted according to the evolving accelerator conditions. In Phase 2, several quenches of QCS superconducting magnets were accompanied by SuperKEKB aborts, coming too late to prevent large energy depositions in the superconducting coils by the rapidly increasing beam losses. A study of the recorded diamond signals in these events showed that in most cases a lower diamond abort-threshold could have triggered an earlier abort of the beams, reducing in this way the energy deposition by beam losses and preventing the corresponding QCS quenches. As a result of this study, the diamond abort thresholds were lowered, still remaining well above the noise levels (Table~\ref{tab:abort_thresholds}, first line). 

\begin{table}
\centering
\caption{
Abort thresholds in Phase 2 (2018) and Phase 3 (starting in 2019), adopted after initial tests and the adjustments preventing QCS quenches. Range 2 and multiplicity ``at least one sum above threshold" have been adopted, except for the first period (range 1, multiplicity ``at least two"). The abort cycle initially had a frequency of $100$~kHz, modified to $400$~kHz since 18 Feb 2020 (section~\ref{subsec:electronics}).
}
\label{tab:abort_thresholds} 
\begin{tabular}{llll}
\hline\noalign{\smallskip}
date            & label       & integr.                  &  threshold         \\
                   &                & time                     &                         \\
\noalign{\smallskip}\hline\noalign{\smallskip}
28 May 2018 & fast         & $1$~ms                & $20$~mrad    \\
                   & slow       & $1$~s                    & $400$~mrad  \\
\hline 
5 Jun 2019 & fast        & $1$~ms                 & $140$~mrad     \\
                   & slow       & $1$~s                    & $4.4$~rad         \\
\hline 
12 Jun 2019 & fast        & $1$~ms                 & $20$~mrad      \\
                   & slow       & $1$~s                    & $10$~rad          \\
\hline
15 Oct 2019 & fast        & $1$~ms                 & $60$~mrad     \\
                   & very fast  & $40$~$\mu$s      & $8$~mrad       \\
\hline
28 Oct 2019 & fast        & $1$~ms                 & $80$~mrad     \\
                   & very fast  & $40$~$\mu$s      & $80$~mrad      \\
\hline
18 Feb 2020 & fast        & $1$~ms                 & $80$~mrad     \\
                   & very fast  & $10$~$\mu$s      & $8$~mrad      \\
\hline\noalign{\smallskip}
\noalign{\smallskip}
\end{tabular}
\end{table}

We also observed that shorter integration time-intervals for the moving sums (section~\ref{subsec:electronics}) allowed to advance the abort request in case of rapidly rising beam losses, with the advantage of minimizing the received radiation dose. Moreover, the reduction of the abort-cycle time in the DCU firmware from $10$ to $2.5$~$\mu$s contributed to the effort of shortening the delays in all the components of the SuperKEKB abort system. Table~\ref{tab:abort_thresholds} shows the evolution of diamond beam abort parameter settings during Phases 2 and 3. 

\subsection{Abort features and statistics}
\label{subsec:abort_features}

\begin{figure}[htbp]
	\centering
	\subfloat[]{{\includegraphics[width=7cm]{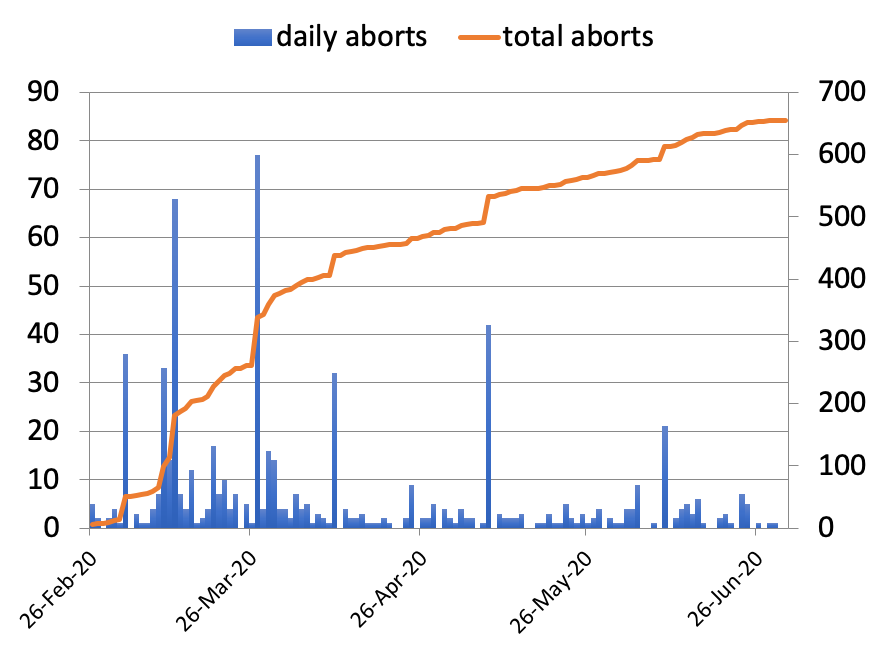} }}%
	\hfill
	\subfloat[``Very fast" abort]{{\includegraphics[width=8cm]{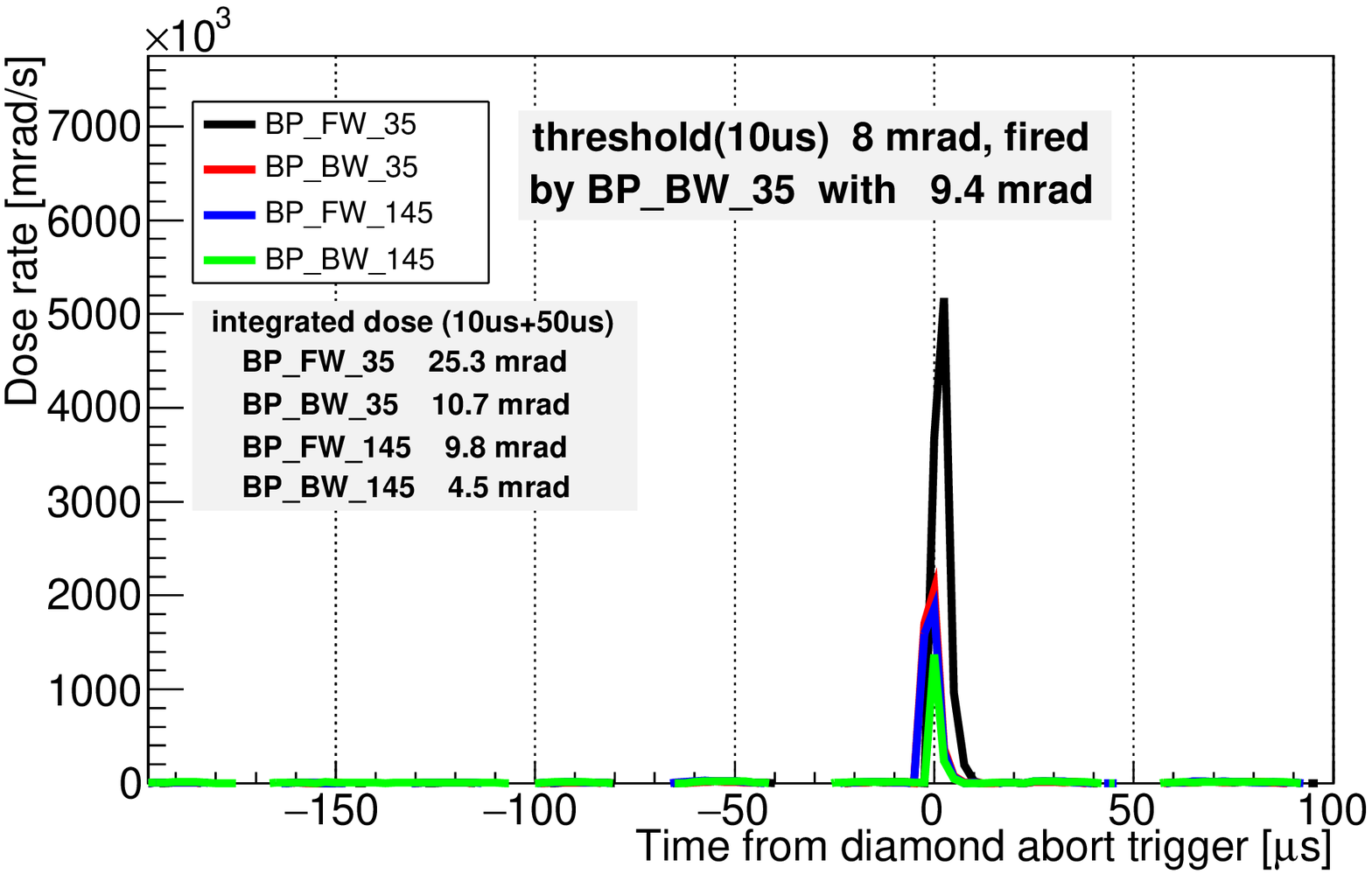} }}%
	\hfill
	\subfloat[``Fast" abort]{{\includegraphics[width=8cm]{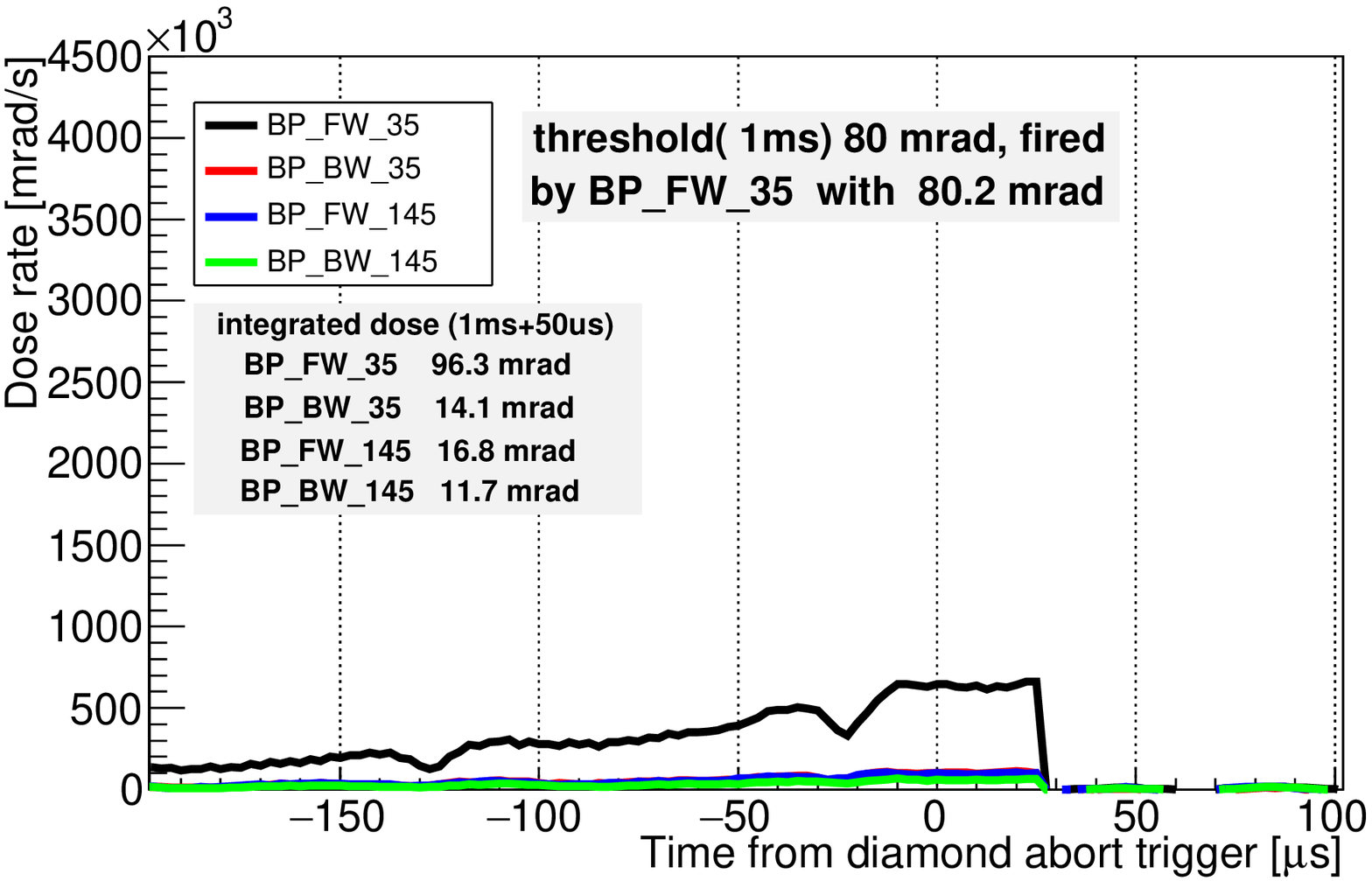} }}
	\caption{(a) Histogram of the number of aborts per day (scale on the left) in February-June 2020, and cumulative number of diamond abort requests (continuous line, scale on the right). (b, c) Two examples of post-abort buffer memory dump: dose rate as a function of time, whose zero corresponds to the reconstructed threshold-crossing time.}
	\label{fig:AbortsStat}
\end{figure}

A total of 305 abort requests were delivered by the diamond-based system in 2019, 655 in 2020 spring runs. Most aborts were determined by the ``fast" threshold initially, and by the ``very fast" threshold after its implementation. The latest ``very fast" threshold setting, with a time window of $10$~$\mu$s for the moving sum (four times $2.5$~$\mu$s), corresponds to an average dose rate of $0.8$~krad/s in the time interval, or to a single $3.2$~krad/s spike concentrated in one $2.5$~$\mu$s sample. 

Figure~\ref{fig:AbortsStat}(a) shows the number of aborts per day in 2020. The higher peaks correspond to accelerator tuning. The decreasing slope of the cumulative distribution indicates a decrease in the average number of aborts per day after the initial period, correlated with improvements in accelerator conditions, namely tighter collimators and stable injections.

Figure~\ref{fig:AbortsStat}(b,c) shows two examples of post-abort memory dump for four aborting diamond detectors, after pedestal subtraction and conversion to dose units. The first event corresponds to a very sharp rise of the dose rate, which reaches a large value of about 5 krad/s immediately after crossing the abort threshold and before the beam abort process is completed. The second event shows a much less frequent occurrence of more slowly rising dose rate, whose integral over $1$~ms crosses the ``fast" threshold of $80$~mrad.

\section{Conclusions and outlook}
\label{conclusions}
The Belle II diamond-based beam-loss monitoring and beam-abort system, designed and assembled at the INFN Trieste laboratories, was installed and operated successfully in the interaction region of the SuperKEKB electron-positron collider. It has been reliably providing dose-rate monitoring data to keep the radiation budget of Belle II inner detectors under control, and has given valuable feedback for accelerator studies and tuning. The participation in the beam-abort system, with the delivery of a substantial fraction of the abort requests, has ensured an effective protection of the Belle II inner detectors and of the SuperKEKB superconducting final-focus magnets.

To meet the challenge of SuperKEKB long-term operation, with the projected future large increase in luminosity, possible improvements of the diamond system are considered. In particular, the performance of the beam-loss monitoring would be improved by electronics with a wider dynamic range and by a more flexible access to dose-rate data with high time resolution in the ring buffer memories. 

\section*{Acknowledgements}
The construction of the diamond-based beam loss monitor and beam abort system was funded by Istituto Nazionale di Fisica Nucleare (INFN) in the framework of the Belle II experiment. The electronics was designed by the Instrumentation and Detectors Laboratory of Elettra Sincrotrone Trieste ScpA. We thank our colleagues of the SVD, PXD, and BEAST groups within the Belle II collaboration for fruitful discussions and comments, and the Machine-Detector Interface group (MDI) of Belle II and SuperKEKB, an essential discussion forum for the operation of the system.



%\section*{References}


\bibliography{mybibfile}

\begin{thebibliography}{10}
\expandafter\ifx\csname url\endcsname\relax
  \def\url#1{\texttt{#1}}\fi
\expandafter\ifx\csname urlprefix\endcsname\relax\def\urlprefix{URL }\fi
\expandafter\ifx\csname href\endcsname\relax
  \def\href#1#2{#2} \def\path#1{#1}\fi

\bibitem{Ohnishi:2013fma}
Y.~Ohnishi, et~al., {Accelerator design at SuperKEKB}, PTEP 2013 (2013) 03A011.
\newblock \href {https://doi.org/10.1093/ptep/pts083}
  {\path{doi:10.1093/ptep/pts083}}.

\bibitem{Abe:2010gxa}
T.~Abe, et~al., {Belle II Technical Design Report }\href
  {http://arxiv.org/abs/1011.0352} {\path{arXiv:1011.0352}}.

\bibitem{Adachi:2018qme}
I.~Adachi, T.~Browder, P.~Kri\v{z}an, S.~Tanaka, Y.~Ushiroda, {Detectors for
  extreme luminosity: Belle II}, Nucl. Instrum. Meth. A 907 (2018) 46--59.
\newblock \href {https://doi.org/10.1016/j.nima.2018.03.068}
  {\path{doi:10.1016/j.nima.2018.03.068}}.

\bibitem{Ikeda:2017xtn}
H.~Ikeda, J.~Flanagan, H.~Fukuma, T.~Furuya, M.~Tobiyama, {Beam Loss and Abort
  Diagnostics during SuperKEKB Phase-I Operation}, in: {5th International Beam
  Instrumentation Conference}, 2017, p. TUAL03.
\newblock \href {https://doi.org/10.18429/JACoW-IBIC2016-TUAL03}
  {\path{doi:10.18429/JACoW-IBIC2016-TUAL03}}.

\bibitem{Lewis:2018ayu}
P.~Lewis, et~al., {First Measurements of Beam Backgrounds at SuperKEKB}, Nucl.
  Instrum. Meth. A 914 (2019) 69--144.
\newblock \href {http://arxiv.org/abs/1802.01366} {\path{arXiv:1802.01366}},
  \href {https://doi.org/10.1016/j.nima.2018.05.071}
  {\path{doi:10.1016/j.nima.2018.05.071}}.

\bibitem{Ohnishi:2020}
Y.~Ohnishi, Y.~Funakoshi, {Highlights from SuperKEKB Commissioning for early
  stage of Nano-Beam Scheme and Crab Waist Scheme}, in: {ICHEP 2020},
  \url{https://indico.cern.ch/event/868940/contributions/3815760/}.

\bibitem{Raimondi:2008zzb}
P.~Raimondi, {Crab waist collisions in DAFNE and Super-B design}, in: {11th
  European Particle Accelerator Conference}, 2008, pp. 1898--1902.

\bibitem{Arinaga:2013pxa}
M.~Arinaga, et~al., {Progress in KEKB beam instrumentation systems}, PTEP 2013
  (2013) 03A007.
\newblock \href {https://doi.org/10.1093/ptep/pts095}
  {\path{doi:10.1093/ptep/pts095}}.

\bibitem{Re:2004rc}
V.~Re, et~al., {Radiation hardness and monitoring of the BABAR vertex tracker},
  Nucl. Instrum. Meth. A518 (2004) 290--294.
\newblock \href {https://doi.org/10.1016/j.nima.2003.11.002}
  {\path{doi:10.1016/j.nima.2003.11.002}}.

\bibitem{Edwards:2005hi}
A.~J. Edwards, M.~Bruinsma, P.~Burchat, H.~Kagan, R.~Kass, D.~P. Kirkby, B.~A.
  Petersen, T.~Pulliam, {Radiation monitoring with CVD diamonds in BaBar},
  Nucl. Instrum. Meth. A552 (2005) 176--182.
\newblock \href {https://doi.org/10.1016/j.nima.2005.06.028}
  {\path{doi:10.1016/j.nima.2005.06.028}}.

\bibitem{Sfyrla:2007ng}
A.~Sfyrla, R.~Eusebi, R.~Tesarek, P.~Dong, C.~Schrupp, R.~Wallny, {Beam
  Condition Monitoring with Diamonds at CDF}, IEEE Trans. Nucl. Sci. 55 (2008)
  328--332.
\newblock \href {https://doi.org/10.1109/RTC.2007.4382739,
  10.1109/TNS.2007.913492} {\path{doi:10.1109/RTC.2007.4382739,
  10.1109/TNS.2007.913492}}.

\bibitem{ref_atlas}
V.~Cindro, et~al., {The ATLAS beam conditions monitor}, JINST 3 (2008) P02004.

\bibitem{ref_cms}
R.~Hall-Wilton, et~al., {Fast beam conditions monitor (BCM1F) for CMS}, in:
  {2008 IEEE Nuclear Science Symposium Conference Record}, 2008, pp.
  3298--3301.
\newblock \href {https://doi.org/10.1109/NSSMIC.2008.4775050}
  {\path{doi:10.1109/NSSMIC.2008.4775050}}.

\bibitem{Bassi:2020xy}
G.~Bassi, et~al., {Calibration of diamond detectors for dosimetry in beam-loss
  monitoring}, to be submitted to Nucl. Instrum. Meth. A, arXiv:2102.03273.

\bibitem{ref_e6}
Element Six (UK) Ltd., \url{http://www.e6.com/}.

\bibitem{ref_cividec}
CIVIDEC Instrumentation GmbH, \url{https://cividec.at/}.

\bibitem{ref_rogers}
Rogers Corporation, \url{http://www.rogerscorp.com/}.

\bibitem{ref_cyclone}
\url{https://www.intel.com/content/www/us/en/products/programmable/fpga/cyclone-v.html}.

\bibitem{ref_ADC}
Analog Devices AD9653,
  \url{https://www.analog.com/media/en/technical-documentation/data-sheets/AD9653.pdf}.

\bibitem{ref_opamp}
Analog Devices LTC6268,
  \url{https://www.analog.com/media/en/technical-documentation/data-sheets/AD9653.pdf}.

\bibitem{ref_epics}
\url{https://epics-controls.org/resources-and-support/base/series-3-14/}.

\bibitem{Bassi:2018mdf}
G.~Bassi, {Radiation monitor with diamond sensors for the Belle II experiment
  at SuperKEKB}, Laurea thesis, Trieste U. (2018), available at
  \url{https://inspirehep.net/literature/1704204}.

\bibitem{ref_SAD}
\url{http://acc-physics.kek.jp/SAD/}.

\bibitem{Agostinelli:2002hh}
S.~Agostinelli, et~al., {GEANT4--a simulation toolkit}, Nucl. Instrum. Meth. A
  506 (2003) 250--303.
\newblock \href {https://doi.org/10.1016/S0168-9002(03)01368-8}
  {\path{doi:10.1016/S0168-9002(03)01368-8}}.

\bibitem{Manfredi:2019lna}
R.~Manfredi, {Diamond-detector commissioning and tracking-efficiency studies in
  the Belle II experiment}, Laurea thesis, Trieste U. (2019), available at
  \url{https://inspirehep.net/files/0b43ecac548eb0f8ba3724cafce35791}.

\bibitem{Mimashi:2014dya}
T.~Mimashi, N.~Iida, M.~Kikuchi, T.~Mori, K.~Abe, A.~Sasagawa, A.~Tokuchi,
  {SuperKEKB Beam abort System}, in: {5th International Particle Accelerator
  Conference}, 2014, p. MOPRO023.
\newblock \href {https://doi.org/10.18429/JACoW-IPAC2014-MOPRO023}
  {\path{doi:10.18429/JACoW-IPAC2014-MOPRO023}}.

\end{thebibliography}

%\begin{thebibliography}{00}
%\end{thebibliography}

%\newpage

%\appendix
%\section{Figures}

%\begin{figure*}[htbp]
%	\begin{center}
%		\includegraphics[width=12cm]{figures/fig_DoseRates.pdf}
%		\caption{An example of display of dose rates from an early stage of accelerator operations, with beam injections every few minutes and relatively small beam currents (LER up to about 120 mA, HER up to about 300 mA). A selection of four diamond detectors are displayed in mrad/s together with the circulating beams currents in mA. On the left side: radiation spikes are correlated with noisy injections; on the right side: two of the diamonds are highly sensitive to the losses of the positron (LER) beam. }
%		\label{fig:DoseRates}
%	\end{center}
%\end{figure*}

\end{document}